\documentclass{article}
\usepackage[a4paper, total={5.5in, 8in}]{geometry}
\usepackage{footmisc}
\RequirePackage[colorlinks=TRUE,allcolors=blue]{hyperref}
\usepackage[caption=false,labelfont=sf]{subfig}
\usepackage[utf8]{inputenc} 
\usepackage{graphicx}
\usepackage{amsmath,amssymb,amsfonts}
\usepackage{array}
\usepackage{textcomp}
\usepackage{xcolor}
\usepackage{url}
\usepackage{comment}
\usepackage{xspace}
\usepackage{booktabs}
\usepackage{algpseudocode} 
\usepackage{graphicx}
\usepackage{natbib}

\algrenewcommand\algorithmicrequire{\textbf{Input:}}
\algrenewcommand\algorithmicensure{\textbf{Output:}}

\newcommand{\Szemeredi}{Szemer\'edi}
\newcommand{\cU}{\mathcal{U}}

\newcommand{\smax}{s_{\rm max}}
\newcommand{\tmax}{t_{\rm max}}
\newcommand{\weq}{\ = \ }

\newcommand{\abs}[1]{{\lvert#1\rvert}}

\newcommand{\ceiling}[1]{\left\lceil #1 \right\rceil}

\DeclareMathOperator*{\argmin}{arg\,min}
\newcommand{\Poi}{\operatorname{Poi}}
\newlength{\wid}
\newlength{\hei}

\begin{document}

\title{A network community detection method with integration of data from multiple layers and node attributes}

\author{Hannu Reittu\footnote{VTT Technical Research Centre, PO Box 1000, 02044, Finland} \and Lasse Leskel\"a\footnote{Aalto University, Department of Mathematics and System Analysis, Otakaari 1, 02150 Espoo, Finland} \and Tomi R\"aty\footnote{Microsoft, One Microsoft Way, Redmond WA 98052, USA}}

\date{10 January 2023}

\maketitle

\begin{abstract}
Multilayer networks are in the focus of the current complex network study. In such networks multiple types of links may exist as well as many attributes for nodes. To fully use multilayer --- and other types of complex networks in applications, the merging of various data with topological information renders a powerful analysis. First, we suggest a simple way of representing  network data in a data matrix where rows correspond to the nodes, and columns correspond to the data items. The number of columns is allowed to be arbitrary, so that the data matrix can be easily expanded by adding columns. The data matrix can be chosen according to targets of the analysis, and may vary a lot from case to case. Next, we partition the rows of the data matrix into communities using a method which allows maximal compression of the data matrix. For compressing a data matrix, we suggest to extend so called regular decomposition method for non-square matrices. We illustrate our method for several types of data matrices, in particular, distance matrices, and matrices obtained by augmenting a distance matrix by a column of node degrees, or by concatenating several distances matrices corresponding to layers of a multilayer network.
We illustrate our method with synthetic power-law graphs and two real networks: an Internet autonomous systems graph and a world airline graph. We compare the outputs of different community recovery methods on these graphs, and discuss how incorporating node degrees as a separate column to the data matrix leads our method to identify community structures well-aligned with tiered hierarchical structures commonly encountered in complex scale-free networks.
\end{abstract}

\section{Introduction}

Networks annotated with node attributes and link attributes form a rich class of data structures. For example, multilayer and multiplex networks are obtained when nodes and links sharing a common attribute are identified as a layer \citep{kivela,frich}. This article presents a simple method for identifying communities in such networks.  The first step is to combine various relevant data sets into a single data matrix, denoted $M$, in which rows correspond to network nodes and columns to data items. The second step is to arrange the rows of $M$ into disjoint groups, called communities, using a regular decomposition (RD) method adopted from \citep{reittujoensuu,reittuBigData2018,reittuspringer,bigpaper}. RD determines communities by a partition of nodes which allows a maximal compression of $M$. This is similar in spirit to nonparametric Bayesian methods associated with stochastic block models \citep{peixoto}. However, in our case we suggest to partition only the rows. This can be seen as an extreme case of block modelling in which every column is considered as a block. To determine the number of communities, we suggest using the Minimum Description Length Principle (MDL) following the RD method \citep[e.g.][]{reittujoensuu,bigpaper}. In this approach, each partition of the node set induces a certain probability distribution on the space of data matrices. The rounded-up integer part of minus logarithm of the probability
of the observed data matrix $M$ is the length of the Shannon code for $M$ \citep[e.g.][]{coverthomas}. Such a coding exists, provably, but there is no need to know how it is constructed. The length of such a code is just used to measure the goodness of fit of a model. 
According to the MDL principle, the full coding length of $M$ is the sum of the Shannon code length and the prefix code lengths of all parameters of the associated probability distribution. For instance, one of such parameters is the number of communities $k$; its approximate code length is $\log k$. By minimising the full code length, MDL is capable of optimising all parameters
\citep[see][]{mdlpeixoto,grunwald,bigpaper}. 
In our sample cases we use graph distances as data items associated to nodes. The use of distance matrix as a basis for spectral community detection was suggested in \citep{bickel}, and as basis for RD in \citep{reittuBigData2018}. One benefit of such a choice is that in a sparse connected network, every pair of nodes has a nonzero distance entry, whereas most entries of the adjacency matrix are zero \citep{reittuBigData2018}.
%
%
In multiplex networks, one approach of constructing a data matrix is to concatenate distance matrices associated with distinct layers so that $M = [D_1 \dots D_m]$ where $D_s$ indicates the distance matrix of layer $s=1,\dots,m$. 
%
%
In the case of directed networks, each distance matrix $D_s$ may be replaced by $[D_s, D^T_s]$ where the $ij$-entry of $D_s$ equals the shortest directed path length from $i$ to $j$, and the transposed matrix, $D^T_s$, gives the corresponding path lengths in the reverse direction.
In the aforementioned cases, the data matrix is determined by the adjacency matrix. However, because our method makes no assumptions on the number of columns of the data matrix, arbitrary type of node attributes can easily be incorporated as auxiliary columns in the data matrix.

The performance of the proposed method is illustrated by analysing three cases, one synthetic network and two real-world networks. First, we consider a synthetic power-law random graph in which each node possesses a capacity characterising the propensity of link formation with other nodes \citep{norrosreittu,remcobook}. These capacities are considered as extra data items forming one column in the data matrix $M$. When the capacities follow a power-law distribution, a nontrivial asymptotic graph structure emerges \citep{reittunorros,soft} where nodes can be grouped into tiers so that nodes with capacity inside a certain interval form a tier, and the tiers characterise shortest path lengths in the network \citep{remcodistance}. Our aim is to identify network communities that can be related to the distribution of the shortest path lengths and consequently to the tiers. Along with high degree variability, another challenge is that the whole tier structure has a vanishing relative size in the large-graph limit. Usual community detection algorithms are prone to ignore such small-scale communities. 

Second, as an example of a single-layer real network, we consider a snapshot of the Internet topology in which the nodes are autonomous systems (AS) and the edges are direct-peering relationships between them
\citep{Gastner_Newman_2006}.
The data matrix $M$ equals the graph distance matrix with an extra column of node degrees added. 

Third, as an example of a multiplex real-life network, we investigate a world airline graph, in which nodes are airports, links are airways, and layers correspond to carriers. This graph is directed and has a skewed degree distribution, with few high-degree nodes acting as hubs. We consider concatenated two-way distance matrices of the layers as the data matrix. In this example, we also demonstrate how to deal with missing data values which correspond to not fully connected layers.

Finally, we compare our method with some other widely used approaches in community detection and data compression.

\subsection{Related work and main new contributions}
\label{sec:RelatedAndContributions}

Community detection is by now a well-developed field having a literature covering lots of efficient computational methods and deep theoretical treatments of consistency \citep{Girvan_Newman_2002,Karrer_Newman_2011,fortunato21,peixoto,yunpeng,Lei_Rinaldo_2015,Zhang_Zhou_2016,Xu_Jog_Loh_2020,Avrachenkov_Dreveton_Leskela_2022,bollabook}. Theoretically, for a given statistical generative model, the most accurate community recovery is achieved by a maximum likelihood estimator \citep{yunpeng,Zhang_Zhou_2016} but implementing this is usually computationally infeasible for large networks. Whereas popular adjacency matrix based spectral clustering methods seek to cluster nodes by their \emph{expansion profiles} \citep{Lei_Rinaldo_2015}, an alternative approach is to cluster nodes by their \emph{distance profiles} \citep{reittuBigData2018}. The relevance of distances in identifying network communities has not yet been much studied empirically. In many cases expansion and distance profiles lead to similar results, but in certain cases distance profiles might expose soft hierarchies which are not easy to detect directly from the adjacency matrix.

In the present article we propose a general approach where nodes are clustered based on generic \emph{data profiles} with an arbitrary number of numerical data associated to every node.  Community recovery in our generalised approach is based on maximising a Poisson likelihood.  This is similar to the stochastic block model (SBM) in that both generate random matrices with rows corresponding to nodes, and the probability distribution of each row is determined by the community of the corresponding node. Nevertheless, there is one crucial difference: whereas SBM generates samples of the full graph (adjacency matrix), our model captures a user-specified set of features associated with each node.  Choosing adjacency indicator variables as features, we obtain the adjacency matrix as a special case.  When fitted to an adjacency matrix, our method becomes similar to a classical SBM-based maximum likelihood estimator.  SBM-based community recovery methods are known to suffer from degree bias which can be avoided by employing a degree-corrected SBM method \citep{Karrer_Newman_2011,yunpeng}.  Instead of adjacency matrices, our model can be fitted into arbitrary node features. Our examples focus on graph distances and node degrees as data items. When fitted to distance data, our model does not implicitly impose Poisson degree distributions, and therefore, our method applied to distance matrices provides an alternative way to avoid degree bias in community detection.


There is no canonical definition of a community in networks. In the present article we interpret communities from an information-theoretical viewpoint of the MDL principle \citep{grunwald}. As a result, an objective measure of the success of community detection is the compression rate of the data at hand. In the current work we suggest to use this method to generic data matrices describing a network. The novelty of our work comes from suggesting a systematic way on finding graph communities which reflect a multitude of data items associated to the network by partitioning the rows of the corresponding data matrix. 
We demonstrate our method using graph distances as a relation between the nodes, augmented with node degrees as scalar data items. We also demonstrate how to deal with a case when relations do not exist between all pairs of nodes, in the case of 
a directed multiplex network.


Several complementary methods exist for identifying communities, say, in multilayer networks. For instance, extending the concept of modularity \citep{modularity} to multilayer setting \citep{wilson} and identifying modularity flows with information-theoretic tools \citep{dedomenico}. There are also many alternatives for extending graph community detection which takes into account data which is not induced from the topology. \citep{newmanclauset,hric} extend SBM in order to take into account node metadata. Community detection in multilayer networks with node attributes has also been proposed in \citep{multilayercom}, yielding promising results in interpreting the communities and using node attribute for predicting unknown links etc. \citep{multilayercom2} develop SBM for multilayer networks in which node attributes are used for enhancing solving network inference problems. Ideas in these publications could be used to enhance our method using more sophisticated treatment of the data items, which we leave as a subject for further study. In an extended survey, a multitude of methods for community detection are presented and evaluated for various use cases \citep{multiplex}. Development of quantum computers may offer new ways of solving hard community detection problems in the future. For instance, solving the modularity maximization problem can be seen as an instance of quadratic binary optimization, which can be solved on so called quantum annealer realized by the D-Wave with around $5$ thousand quantum bits, \citep{mniszewski}. Another idea is to use \Szemeredi's regularity lemma \citep{Szeme76,tao} for obtaining a quadratic binary cost function, minimum of which yields graph communities, \citep{panning}.

\section{Regular decomposition}

In this section, we describe our method of analysing a network based on a generic data matrix, which describes the network.  Regular decomposition was originally developed in \citep{nepusz2008,reittuetall,pehkonenreittu,reittujoensuu,reittubazsonorros,reittuBigData2018,reittuspringer,bigpaper}. Regular decomposition is inspired by \Szemeredi's regularity lemma \citep{Szeme76}, information theory \citep{grunwald} and stochastic block models \citep{abbe}. In the publication \citep{reittuBigData2018} the regular decomposition method  was used for community detection with a single graph distance matrix as a data matrix. In \citep{haryo} the authors evaluated performance of the regular decomposition method for a generic data clustering. 
In this work, we extend such methods by exploiting more general, non-rectangular,  data matrices as a basis for community detection. In this way we get a flexible method that can find communities that highlight various properties of the network, like degree distribution and distances in multiplex networks and could be used in other cases that can be formulated in a similar way. In the next subsection we expose such a method in more details following and adapting some ideas in the cited works.

\subsection{Data matrix and partition matrix}

Consider a set of nodes indexed by $[n] = \{1,\dots,n\}$, and an $n$-by-$m$ data matrix $M$ in which row $i$ represents data associated with node $i$, and entry $M_{ij}$ represents the value of the $j$\textit{-th} data item associated with node $i$. In a basic setting $M$ equals the adjacency matrix $A$ (with $m=n$) of a single graph, and the rows correspond to \emph{adjacency profiles} of the nodes. Alternatively, network data could be summarised by a distance matrix $D$ (with $m=n$), in which case the rows of the data matrix correspond to \emph{distance profiles}. Indeed, matrices $D$ and $A$ provide equivalent representations of the network topology\footnote{The shortest-path distance matrix $D$ for a graph can be computed from the adjacency matrix $A$ using a standard algorithm, e.g.\ breadth-first search. Conversely, the adjacency matrix $A$ may be recovered from the distance matrix $D$ by noting that $A_{ij}=1$ if and only if $D_{ij}=1$. 
}.
In this article we take a more general approach and allow the number $m$ of data items associated with a node to be arbitrary. This flexibility allows to model multilayer networks by concatenating, say, several adjacency or distance matrices side by side into a single data matrix. Furthermore, any data associated with nodes can easily be concatenated to the data matrix as extra columns. In this more general case, each row of $M$ corresponds to a \emph{data profile} of a node.

Using the data matrix, we partition the node set into $k$ disjoint sets called \emph{communities}. Such a partition can be represented as an $n$-by-$k$ \emph{partition matrix} $R$ with entries
\[
 R_{iu}
 \weq
 \begin{cases}
  1 &\quad \text{if node $i$ is in community $u$},  \\
  0 &\quad \text{else}.
 \end{cases}
\]
In applications some entries of the data matrix $M$ may be undefined or unobserved. These situations are handled by equipping $M$ with an $n$-by-$m$ indicator matrix $C$ in which $C_{ij} =1$ indicated a valid entry, and $C_{ij}=0$ indicates an undefined or unobserved entry. The corresponding matrix elements of $M$, which are not defined or are missing, are replaced by dummy values, which is chosen to be $0$ in all our sample cases. We demonstrate how this works in Section~\ref{sec:AirlineNetwork}.

Our aim is to group nodes into few communities in such a way that description length of $M$ is minimised based on a probabilistic model for the matrix elements of $M$. In other words, we try to find an optimal number of communities $k$ and corresponding  optimal partitioning $\cU_k=\{U_1,U_2,\dots,U_k\}$ of $[n]$ to achieve this. In case of non negative integer-valued $M$, rows in a community $s \in [k]$ are modelled by a sequence of $m$ independent Poisson-distributed random variables denoted as $(X^s_1,X^s_2,\cdots,X^s_m)$. As a result the probabilistic model constitutes of $n \times m$ independent Poisson random variables with $k \times m$ parameters, the expectations of the corresponding variables.
Such a choice is adopted from \citep{reittujoensuu}. Communities are selected in order to minimise the magnitude
\[
 L(\cU_k)
 := - \sum_{s \in [k]} \sum_{d\in U_s } \sum_{j \in [m]}
 \log_2(\mathbb{P}(X^s_j = M_{dj}).
\]
According to classical information theory \citep[e.g.][]{coverthomas} there exists a binary code, the Shannon code, for encoding $M$ with code length $=\ceiling{L}$. We use $L$ as a cost function of the partition $\cU_k$. 
The quality of the found communities is the compression ratio
\[ 
 r(\cU_k)=\frac{L(M)}{L(\cU_k)},
\]
in which $L(M)=\sum_{i,j: M_{ij}>0} \log M_{ij}$ is the number of bits needed to represent data matrix $M$ as a string of integers.

\subsection{Likelihood function}
\label{sec:Poisson}

We employ a statistical latent-variable model in which all observable entries (those with $C_{ji}=1$) of the data matrix $M$ are conditionally independent and Poisson-distributed random variables given the community structure. Furthermore, all rows of $M$ corresponding to the same community are identically distributed. This statistical model is parameterised by an $n$-by-$k$ partition matrix $R$ and an $m$-by-$k$ expectation matrix $\Lambda$ of Poisson variables, and corresponds to likelihood function
\begin{equation}
 \label{eq:Model}
 f(M | \Lambda,R)
 \weq \prod_{j=1}^{n} \prod_{i=1}^{m} \prod_{u=1}^k
 \Poi(M_{ji} | \Lambda_{i u})^{C_{ji} R_{ju}},
\end{equation}
where $\Poi(x | \lambda) = e^{-\lambda} \frac{\lambda^x}{x!}$ is the probability mass function of a Poisson distribution with mean $\lambda$. The corresponding log-likelihood can be written as
\[
 \label{log}
 \log f(M | \Lambda,R)
 \weq \sum_{i=1}^m \sum_{j=1}^n \sum_{u=1}^k R_{ju} C_{ji}
 \Big( M_{ji} \log \Lambda_{iu} - \Lambda_{iu} \Big)
 - \text{const}(M),
\]
where $\text{const}(M) = \sum_{j=1}^n \sum_{i=1}^m C_{ji} \log( M_{ji}!)$ does not depend on the model parameters and can be ignored. The above model is structurally similar to the stochastic block model in which the data matrix has $m=n$ columns and corresponds to the adjacency matrix, and the Poisson distributions are replaced by Bernoulli distributions \citep{Holland_Laskey_Leinhardt_1983,Zhang_Zhou_2016}. In contrast to stochastic block models, the above model allows more flexibility in choosing data matrices with an arbitrary number of columns $m$. Note also that only the rows are grouped in blocks, all columns are treated as separate. In this sense the model has maximal number of variables with respect to the number of columns.

\subsection{Maximum likelihood estimation}
\label{sec:MLE}

Having observed a data matrix $M$, maximum likelihood estimation searches for a partition matrix $R$ of $[n]$ for which the function in Eq.~\ref{log} is maximised.
For any fixed $R$, the $\Lambda$-parameters are set equal to 
\begin{equation}
 \label{eq:MLELambda2}
 \hat\Lambda_{iv}(R)
 \weq \frac{\sum_{j=1}^n M_{ji}R_{jv}C_{ji}}{\sum_{j=1}^n R_{jv}C_{ji}},
\end{equation}
which is the observed average of the $i$\textit{-th} data item in community~$v$. As a consequence, a maximum likelihood estimate of $R$ is obtained by maximising the profile log-likelihood $f(M \, | \, \hat\Lambda(R), R)$, or equivalently minimising
\begin{equation}
 \label{eq:NegLogLikelihood}
 L(R)
 \weq
 \sum_{i=1}^m \sum_{j=1}^n \sum_{v=1}^k R_{jv}C_{ji} \left( \hat \Lambda_{iv}(R) - M_{ji} \log \hat \Lambda_{iv}(R) \right) 
\end{equation}
subject to $n$-by-$k$ partition matrices $R$, in which $\hat \Lambda_{iv}(R)$ is given by \eqref{eq:MLELambda2}. 
We note that the above function can be written as $L(R) = \sum_{j=1}^n \ell_{j Z_j}(R)$, in which
\begin{equation}
 \label{eq:LikelihoodComp}
 \ell_{jv}(R)
 \weq \sum_{i=1}^m C_{ji}\left( \hat \Lambda_{iv}(R) - M_{ji} \log \hat \Lambda_{iv}(R) \right)
\end{equation}
is a normalised minus log-likelihood of the data vector of node $j$, given that $j$ is placed in community $v$ and the rate parameters are equal to $\hat\Lambda(R)$. We note that $L(R)$, up to an additive constant, equal to the description length of the data matrix $M$ in the sense of Shannon coding.

The same algorithm can be used also in case of data matrix with positive real values, which was already shown in \citep{reittujoensuu}. In this case, each matrix element $M_{ij}$ is treated as a parameter (the expectation) of a Poisson distribution. The $\Lambda$-matrix is computed according to Eq.~\ref{eq:MLELambda2} for each partition $R$. Eq.~\ref{eq:NegLogLikelihood} also remains intact, and $L$ equals to Kullback--Leibler divergence between the corresponding Poisson distributions, the original with parameters from data matrix $M$ and those with parameters from Eq. \ref{eq:MLELambda2}. The task is to minimise $L$ which can be done with the same algorithm as in the integer case.  


\subsection{Regular decomposition algorithm}

\begin{figure}[!htbp]
\hspace{2mm}
\includegraphics[width=80mm, keepaspectratio=true]{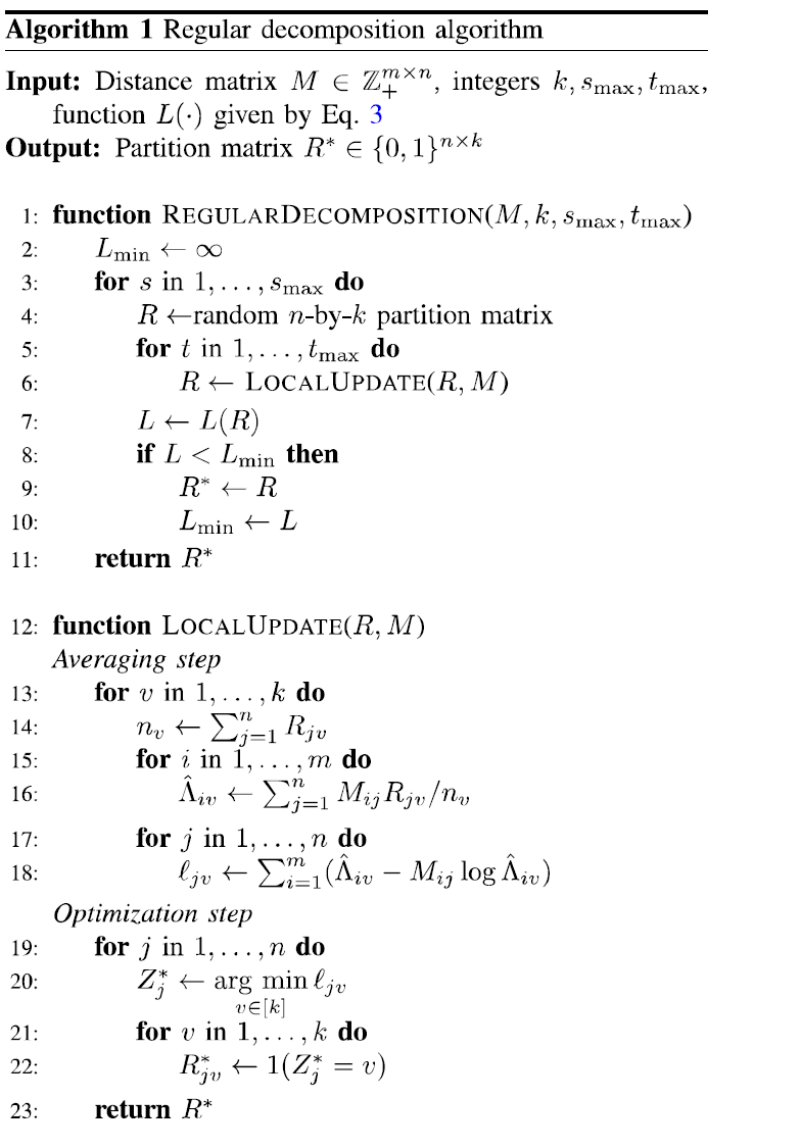}
\vspace{8mm}
\caption{\label{alg:RD} Pseudo code for the Regular decomposition algorithm according to \citet{reittuBigData2018}.}
\end{figure}

Minimising the cost function in Eq.~\ref{eq:NegLogLikelihood} with respect to $R$
is a hard nonlinear discrete optimisation problem with an exponentially large input space of the order of $\Theta(k^n)$, making exhaustive search computationally infeasible. This is why we suggest solving the problem using a greedy Algorithm~\ref{alg:RD} which is an EM (expectation maximisation) type algorithm which alternates between updating $\Lambda$ according to Eq.~\ref{eq:MLELambda2} for a fixed $R$ (E-step) and updating the partition $R$ by greedily updating the community of each node, one by one, to minimise $\ell_{jv}(R)$ in Eq.~\ref{eq:LikelihoodComp} for a fixed $\hat\Lambda$ (M-step). Starting from a uniformly random initial assignment $R_0$, the algorithm finds a local optimum as a limit of the greedy algorithm. Running the greedy algorithm for several initial random states $R_0$, and selecting the community assignment with smallest cost in 
Eq.~\ref{eq:NegLogLikelihood}  as the final output.

The runtime of Algorithm~\ref{alg:RD} is $O(\smax \tmax k (n+m))$, where $\smax$ is the number of random initialisations and $\tmax$ is the number of iterations per optimisation round. Especially, the runtime is linear in $n$ and $m$ for bounded $k,\smax,\tmax$, and is hence scalable for large data sets.

\subsection{Boosted regular decomposition}
\label{sec:Boosted}

Setting up the data matrix for Algorithm~\ref{alg:RD} may require costly preprocessing, e.g.\ computing distances between node pairs (see Sec.~\ref{sec:Distances}). The matrix $M$ may also be simply too large to be treated as a whole, a situation frequently encountered in the realm of so-called big data. 
In such cases the algorithm can be boosted by replacing the data matrix $M$ by a submatrix $M_{VW}$ with a row set $V \subset [n]$ of size $n_0$ and a column set $W \subset [m]$ of size $m_0$, and running Algorithm~\ref{alg:RD} with the submatrix $M_{VW}$ as input.
This results in a partition matrix $R^*$ of the node set $V$. A community assignment of the remaining node set is then computed in a subsequent classification phase where the community index of each node $j \in [n] \setminus V$ is chosen to be $v(j) \in [k]$:
\begin{eqnarray}
 v(j)=\argmin_{v'\in [k]}\ell_{jv'}(R^*),\quad
\ell_{jv'}(R^*) \weq \sum_{i \in W} C_{ji}\left(
 \hat\Lambda_{iv'}(R^*) - M_{ji} \log \hat \Lambda_{iv'}(R^*) \right),
\end{eqnarray}
where
\[
 \hat\Lambda_{iv}(R^*)
 \weq \frac{\sum_{j \in V} M_{ji}R_{jv}^*C_{ji}}
 {\sum_{j \in V} R_{jv}^*C_{ji}}
\]
is the observed average value of data item $i \in W$ among the nodes of $V$ classified into community $v \in [k]$ according to $R^*$.

The runtime of Algorithm~\ref{alg:RD} applied to the submatrix $M_{VW}$ is $O(\smax \tmax k (n_0+m_0))$, and the runtime of the subsequent classification phase is $O(km_0 n)$. Hence, the boosted regular decomposition algorithm has complexity 
$O(\smax \tmax k (n_0+m_0) + k m_0 n)$.
%
%
The feasibility of this boosting approach requires that the row set $V$ is large enough to contain nodes from all communities, and the column set $W$ is a sufficiently informative collection of data items. A simple way of selecting $V$ and $W$ is by random sampling. This approach was developed in \citep{reittuBigData2018,reittuspringer} in which sufficient sample sizes were estimated and convergence proved in some model cases.

\subsection{Estimating the number of communities}
\label{sec:CommunityCount}

Algorithm \ref{alg:RD} requires the number of communities $k$ as an input parameter. However, in most situations this parameter is not a priori known and needs to be estimated from the observed data.  The problem of estimating the number of communities can be approached by recasting the maximum likelihood problem in terms of the minimum description length (MDL) principle \citep{Rissanen_1983,grunwald} where the goal is to select a model which allows a minimum coding length for both the data and the model. MDL adheres to the principle of Occam's razor in which the best hypothesis follows the best compression of data, hence justifying the selection of MDL for this task.

When restricting to the model described in Sec.~\ref{sec:Poisson}, then the $R$-dependent part of the coding length equals $L(R)$ given by \eqref{eq:NegLogLikelihood}, and an  MDL-optimal partition $R^*$ for a given $k$ corresponds to the minimal coding length
$$
 R^* = \arg\min_{R} L(R).
$$
It is not hard to see that $L(R^*)$ is monotonously decreasing as a function of $k$, and a balancing term, the model complexity, is added to select the model that best explains the observed data. The model complexity is the length of a code that uniquely describes the mode itself. In all of our experiments, $L(R^*)$ (the negative log-likelihood) as a function of $k$ becomes essentially a constant above some value $k^*$. Such an elbow point $k^*$ is used as an estimate of $k$ in the experiments in this article, see also \citep{elbow}. In general it might be necessary to have a more sophisticated method using a model complexity term \citep{reittubazsonorros,bigpaper,mdlpeixoto}. 
However, in examples we are using it suffices to use a simplified version of the MDL principle based on the elbow point.

\subsection{Using distances as data items in multiplex networks}
\label{sec:Distances}

In a multiplex network consisting of $s$ directed graphs with a common node set, each graph represents one layer.
The distance matrices of the layers are denoted by $D_1, \cdots, D_s$. If layer $r$ contains no path from node $i$ to node $j$, we declare the corresponding entry as missing by setting $(C_r)_{ij} = 0$, and we may define $(D_r)_{ij} = 0$ without loss of generality. As a result, we obtain $s$ indicator matrices $C_1, \dots, C_s$.
When layers are undirected, data about path lengths is encoded in a concatenated data matrix
\begin{equation}
 \label{eq:MultilayerUndirected}
 M \weq [D_1, D_2, \dots, D_s],
\end{equation}
and the corresponding indicator matrix is $C = [C_1, C_2, \dots, C_s]$. In case of directed layers,
data about directed path lengths is encoded in matrix
\begin{equation}
 \label{eq:MultilayerDirected}
 M \weq [D_1, D_1^T, D_2, D_2^T, \dots, D_s, D_s^T],
\end{equation}
where row $i$ of matrix $D_r$ (resp.\ $D_r^T$) contains the shortest directed path lengths from $i$ to other nodes (resp.\ from other nodes to $i$). The corresponding indicator matrix is denoted
$C = [C_1, C_1^T, C_2, C_2^T, \dots, C_s, C_s^T].$
The data matrix $M$ and the indicator matrix $C$ are then given as input to Algorithm~\ref{alg:RD}, and the optimal number of communities is determined as in Sec.~\ref{sec:CommunityCount}.


Computing the distance matrix in an unweighted directed graph with $n$ nodes and $e$ links has complexity $O(n(n+e))$ using breadth-first search \citep{BangJensen_Gutin_2009}.
The regular decomposition algorithm based on distances can be boosted by computing distances only for a restricted set of reference nodes $W \subset [n]$ of size $m_0$, resulting in an $n$-by-$m_0$ distance matrix $D$ in which $D_{ij}$ equals the distance from node $i \in [n]$ to node $j \in W$, and $D$ can be computed using breadth-first search in $O(m_0(n+e))$ time. The same complexity bound is valid for concatenated data matrices of form \eqref{eq:MultilayerUndirected}--\eqref{eq:MultilayerDirected} in multilayer networks with a bounded number of layers. Then we may apply the boosted regular decomposition algorithm (Sec.~\ref{sec:Boosted}) with $V=[n]$ and $W$ as above. The total complexity of the preprocessing step (restricted distance matrix computations) and boosted regular decomposition algorithm then equals $O(\smax \tmax k (n+m_0) + k m_0 n + m_0(n+e))$. For bounded number of communities $k$ and bounded iteration parameters $\smax,\tmax$, this bound is $O(m_0(n+e))$.  
Especially, by selecting $m_0$ constant, we obtain a scalable algorithm for sparse massive networks with $e=O(n)$, capable of identifying communities in linear time with respect to the number of nodes $n$.

\section{Power-law graphs}

Most real networks are inhomogeneous. In particular, this is true for graphs where nodes posses features that correlate with graph topology. Furthermore, sparsity is commonplace, because links are expensive to maintain.
Many real networks have highly varying degrees, with most nodes having a small number of neighbours, and very few nodes having a huge number of neighbours as was pointed out in \citep{barabasi}. The high-degree nodes usually play an important role as hubs in the network. Already two decades ago, a highly influential study by \citep{3faloutsos} revealed that the Internet has this kind of topology.  

In this section we summarise a simple generative model for sparse random graphs with a power-law degree distribution (Sec.~\ref{sec:PowerLawGraph}), and apply the regular decomposition algorithm to a synthetic graph sampled from the model, first using distances (Sec.~\ref{sec:PowerLawDistances}), and then using distances and degrees (Sec.~\ref{sec:PowerLawDistancesDegrees}). Our aim is to show that the used data matrix has a profound effect on the community structure.

\subsection{Poissonian power-law graph}
\label{sec:PowerLawGraph}


A simple random graph model can be induced from a random graph process described in \citep{norrosreittu,remcobook} which we call a Poissonian power-law graph. We first sample node attributes $\lambda_1, \dots, \lambda_n$ independently at random from a probability distribution on the nonnegative reals, and thereafter connect each unordered node pair $ij$ by a link with probability $1-\exp\{-\lambda_i\lambda_j/\lambda\}$, independently of other node pairs, where $\lambda = \sum_i \lambda_i$. When the node attributes are distributed according to a power-law  with density exponent $\tau \in (2,3)$, we obtain a generator of a sparse random graph where the degree distribution has a finite mean and infinite variance.

In such power-law graphs, quite a rich topological structure spontaneously arises in the limit of large graph size and with probability tending to one. The nodes are categorised into sets called \emph{tiers} according to their degree, such groping we call 'soft-hierarchy' \citep{norrosreittu,reittunorros,soft,vulnerab}.
The top tier $V_0$ is formed, asymptotically as $n\rightarrow\infty$, by nodes with degrees in the range $(n^{1/2}, \infty)$, and the other tiers $V_k$ are formed by nodes with degrees in range $(n^{\beta_k}, n^{\beta_{k-1}}]$, in which $\beta_0 = 1/2$ and $\beta_k=(\tau-2)^k/(\tau-1)+o(1)$ for $k \ge 1$.
For large values of $n$, it is known that with high probability, the top tier $V_0$ is fully connected, and further, every node in $V_k$ has a link to $V_{k-1}$ for all $k$ up to order $\log\log n$
\citep{norrosreittu,remcodistance,globecom}.
The subgraph induced by the union of the tiers $V_k$ for $0 \le k \le \log \log n$ is called the \emph{core network}. Most of the nodes have small degrees and, as a result, are outside the core. However, any node is at a very short distance, compared to $\log\log n$, from the bottom tier. This explains ultra-short distances in the graph. 

Our aim is to show, through experiments on synthetic and real data, that our version of the regular decomposition algorithm can identify soft hierarchical structures by using network distances and node degrees together as a data matrix. The soft hierarchy is a compact description of the organisation of shortest paths between most of the nodes in power-law graphs of the type we are interested in. Regular decomposition compresses the distance matrix, and that is why it is likely that a soft hierarchy shows up in the resulting short description of the matrix. Degrees are also essential in describing the soft hierarchy, and that is why their inclusion in data matrix should help. This is why we argue that a degree-augmented and distance-based community structure matches qualitatively with a kind of rough soft hierarchy in synthetic and real power-law graphs. Notably, in expectation $\mathbb{E}\abs{\cup_k V_k}/n\rightarrow 0$, see \citep{norrosreittu,globecom} as $n \to \infty$, which means that communities associated with the layers are very small and which means that such communities are undetectable for typical community-detection algorithms assuming comparable community sizes.

This illustrates how our method should be used. At first there should be an intuitive understanding which data is essential for the problem to be solved. In the current example, the problem is to find the soft hierarchy as a community structure. In other problems the data matrix could be completely different.

\subsection{Regular decomposition using distances}
\label{sec:PowerLawDistances}

\begin{figure}[!htbp]
\centering
\setlength{\hei}{40mm}
\subfloat[]{\includegraphics[height=\hei]{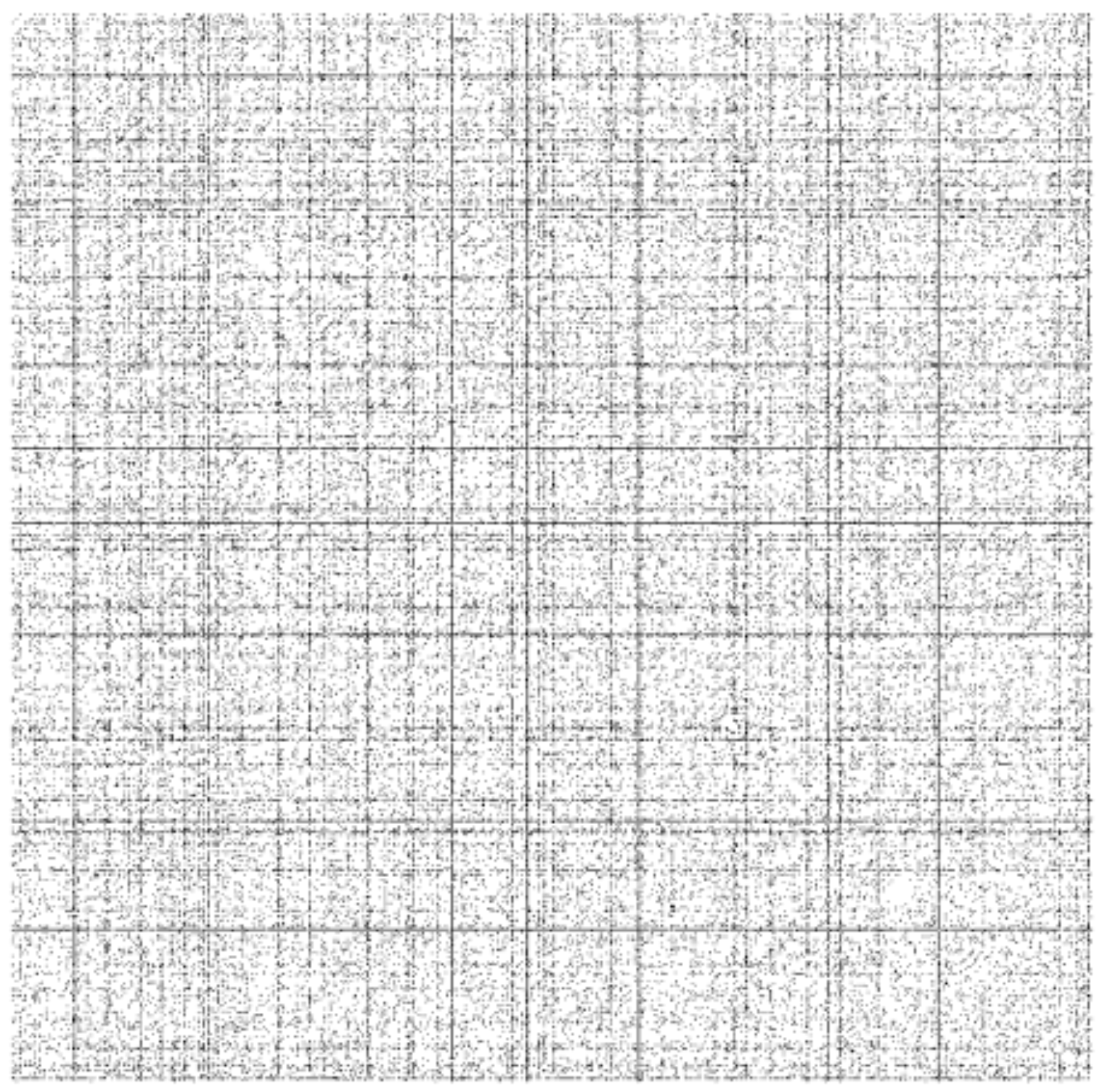}} \hspace{3mm}
\subfloat[]{\includegraphics[height=\hei]{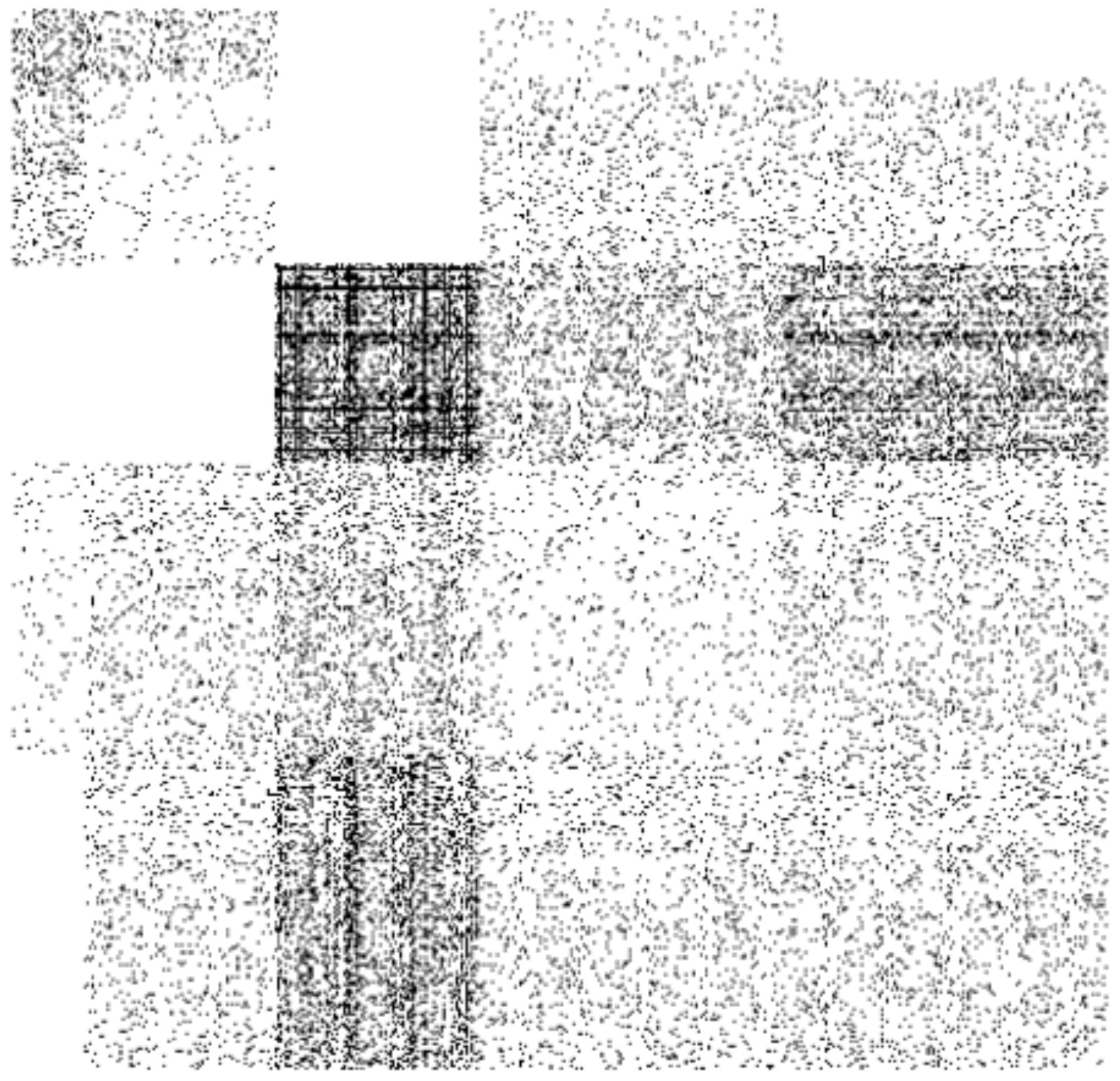}} \hspace{3mm}
\subfloat[]{\includegraphics[height=\hei]{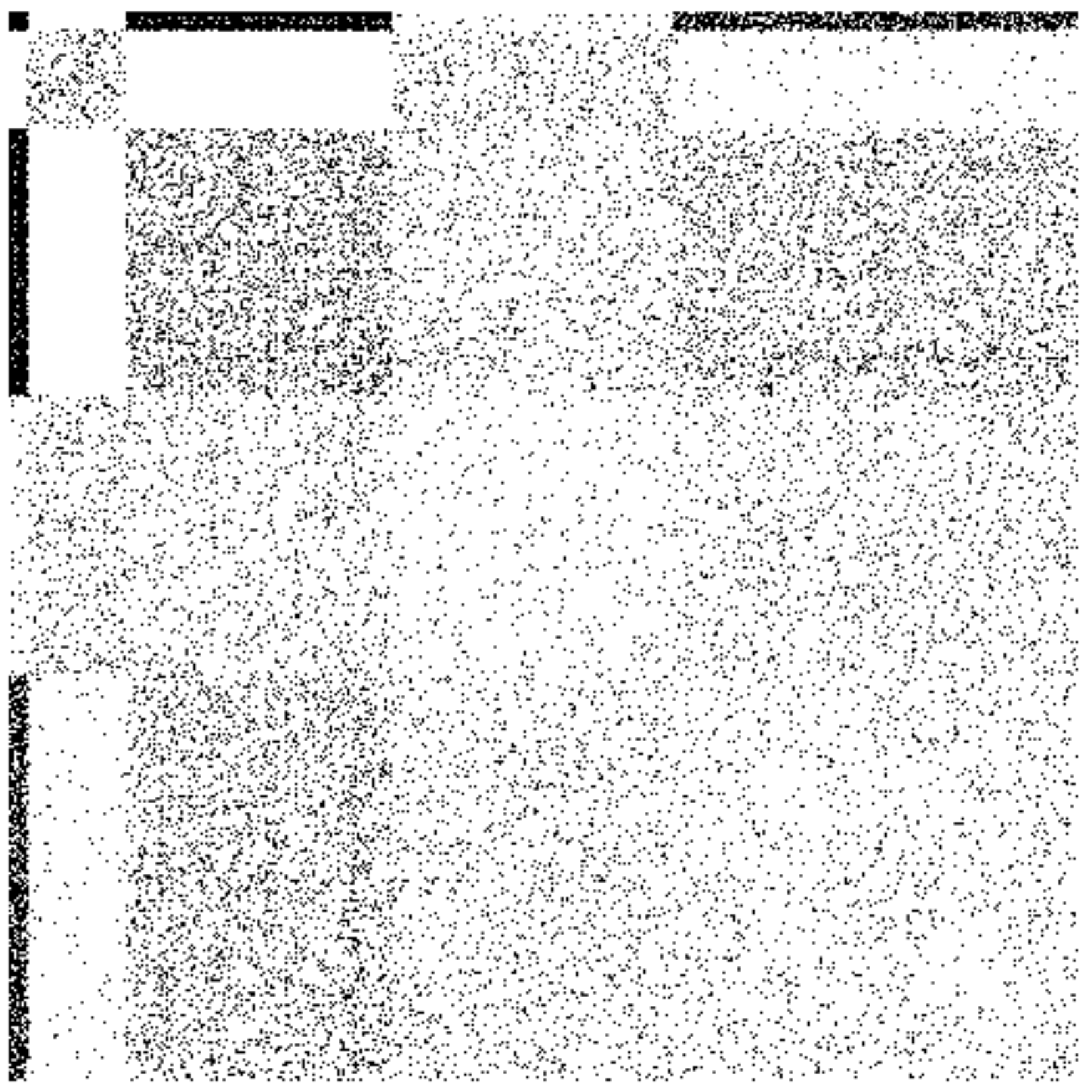}}
\vspace{11mm}
\caption{\label{pwladj} The adjacency matrix of a synthetic power-law graph ($n=7775$) with rows and columns organised according to: (a) random order of nodes, (b) communities identified by distance matrix, (c) communities identified by distance matrix and degrees (right).}
\end{figure}

We generated a power-law graph with ten thousand nodes using the model in Sec.~\ref{sec:PowerLawGraph} with node attributes drawn from a power-law distribution with density exponent $\tau=2.5$, and we extracted its largest connected component as input for subsequent analysis.
As a result, we have a graph with $n=7775$ nodes and adjacency matrix shown in Fig.~\ref{pwladj}:(a). Next we computed the distance matrix of this graph. We identified communities of the graph by applying Algorithm~\ref{alg:RD} using the distance matrix as data matrix.  By experimenting with different values of the number of communities $k$, we found that the cost function in Sec.~\ref{sec:CommunityCount} saturates at $k=5$. This value is identified as the most informative number of communities.

The block structure of the adjacency matrix induced by the identified communities is shown in Fig.~\ref{pwladj}:(b). A clear block structure is revealed with one large block with relatively high density. All five identified communities are rather large. Hence the identified community structure differs remarkably from the theoretical tier structure (Sec.~\ref{sec:PowerLawGraph}) where the top layers are dense and small. This is a natural consequence of partitioning the graph using distance profiles because the small-degree neighbours of high-degree nodes are likely to have similar distance profiles with each other.

\begin{figure}[!htbp]
\centering
\setlength{\wid}{.33\textwidth}
\subfloat[]{\includegraphics[width=\wid]{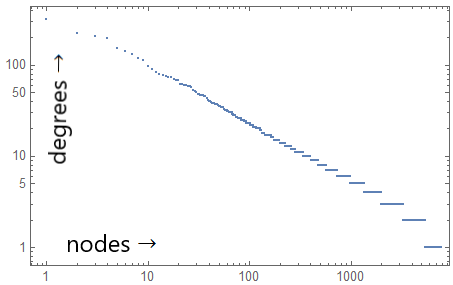}}%
\subfloat[]{\includegraphics[width=\wid]{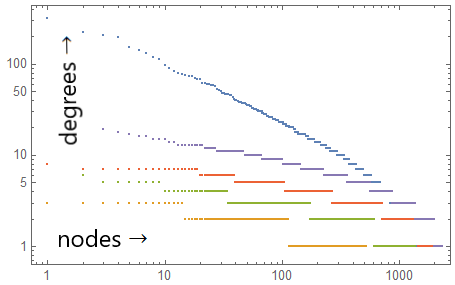}}%
\subfloat[]{\includegraphics[width=\wid]{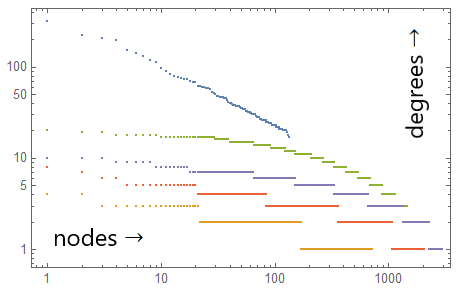}}%
\vspace{11mm}
\caption{\label{pwldeg}
Degrees of nodes in a synthetic power-law graph ($n=7775$) sorted from largest to smallest within each community (indicated by colour). 
(a) Full graph viewed as one community. (b) Nodes organised into communities identified by distance matrix. (c) Nodes organised into communities identified by distance matrix and degrees.
}
\end{figure}

\begin{figure}[!htbp]
\centering
\subfloat[]{\raisebox{5.5mm}{\includegraphics[width=60mm]{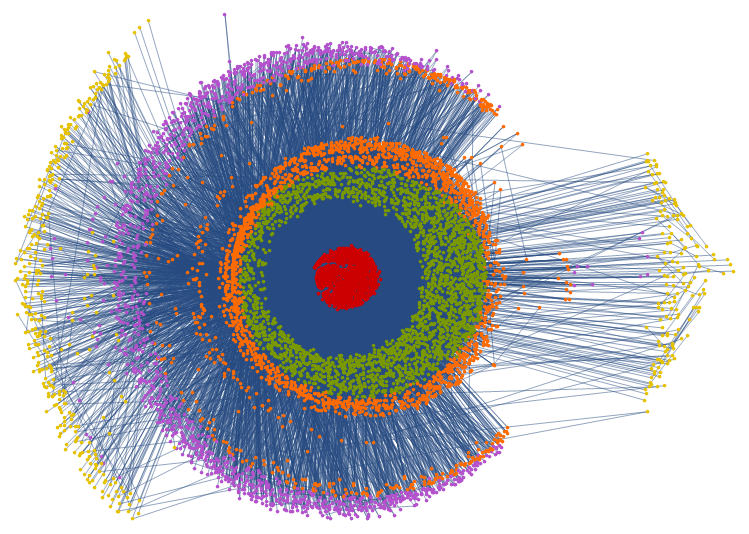}}} \hspace{5mm}
\subfloat[]{\scalebox{-1}[1]{\includegraphics[width=60mm]{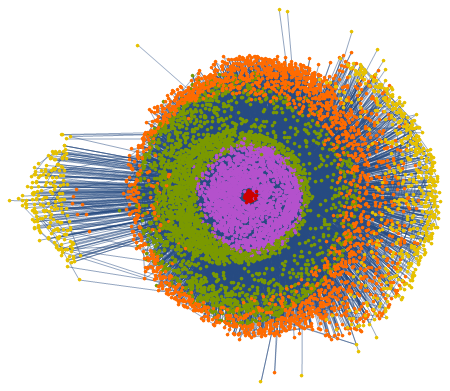}}}
\vspace{8mm}
\caption{\label{pwlcom} Topology of a synthetic power-law graph ($n=7775$) coloured according to the community structure identified by the regular decomposition algorithm with: (a) distance matrix, (b) distance matrix and degrees. }
\end{figure}

The degrees of the graph are plotted in the Fig.~\ref{pwldeg}:(a). The linear shape in the log-log plot on the left is typical for power-law graphs. The degrees of nodes in the five identified communities are shown in Fig.~\ref{pwldeg}:(b). We see that all high-degree nodes are in the same community, but this community also contains nodes of smaller degree, and is quite large. Furthermore, Fig.~\ref{pwlcom}:(a) displays the graph with the five identified communities, plotted using Mathematica's CommunityGraphPlot tool. We see that nodes in the red community are central for connecting nodes in the graph. The ringlike communities around the center appear to roughly play a role similar to tiers in a soft hierarchy, in that most shortest paths between peripheral nodes use the rings to reach the red center.

\subsection{Regular decomposition using distances and degrees}
\label{sec:PowerLawDistancesDegrees}

We continue experimenting with partitioning the same graph sample as in the previous section. Instead of using the distance matrix with 7775 columns as in Sec.~\ref{sec:PowerLawDistances}, we will now use a data matrix with only 101 columns, consisting of 100 randomly sampled columns of the distance matrix and 1 additional column containing the node degrees. The aim of this experiment is twofold. First, we wish to investigate how adding the degrees to the data matrix affects the inferred community structure. Second, we will demonstrate that computing distances to a relatively small set of reference nodes suffices to well characterise the distance profiles of most nodes. 

The adjacency matrix organised according to five identified communities using the modified data matrix is shown in Fig.~\ref{pwladj}:(c). The main difference from the previous case shown in the middle of the same plot, is a small central community with high-degree nodes only. This can be seen in Fig.~\ref{pwldeg}:(c) which presents the node degrees grouped by communities. The blue community contains all high-degree nodes, and its degree sequence does not overlap with other communities. Nodes in the blue community may hence be thought as tier-1 nodes. As a result, the community structure is qualitatively closer to the theoretical tiers of the power-law graph. According to the theory, there are only $\Theta(\log\log n)$ tiers, we may expect only a couple of layers in our sample with $\log\log n = 2.19$.  

Fig.~\ref{pwlcom}:(b) visualises the identified communities from a topological point of view. The small and dense red community in the center corresponds qualitatively to the top tier of the theoretical power-law graph structure in Sec.~\ref{sec:PowerLawGraph}. The second and third largest communities can be seen as tier-2 and tier-3 communities, and the remaining communities form the periphery of low-degree nodes. Incorporating degrees in the data matrix can hence substantially change the community structure, and in our case align the communities better with a soft hierarchy of nodes.

\section{Internet autonomous systems graph}
\label{sec:AS}

\begin{figure}[!htbp]
\centering
\setlength{\hei}{45mm}
\subfloat[]{\includegraphics[height=\hei]{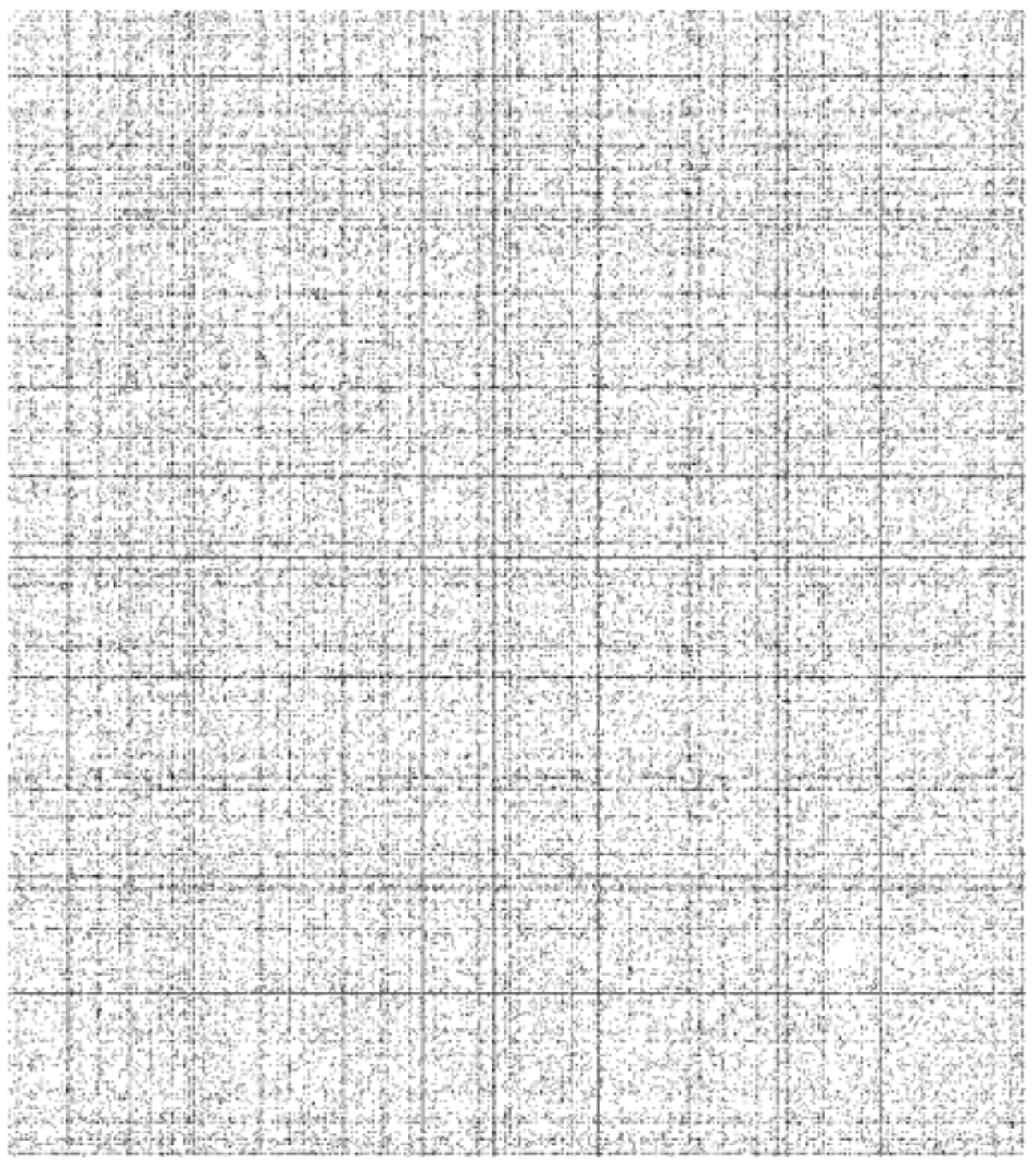}} \hspace{5mm}
\subfloat[]{\includegraphics[height=\hei]{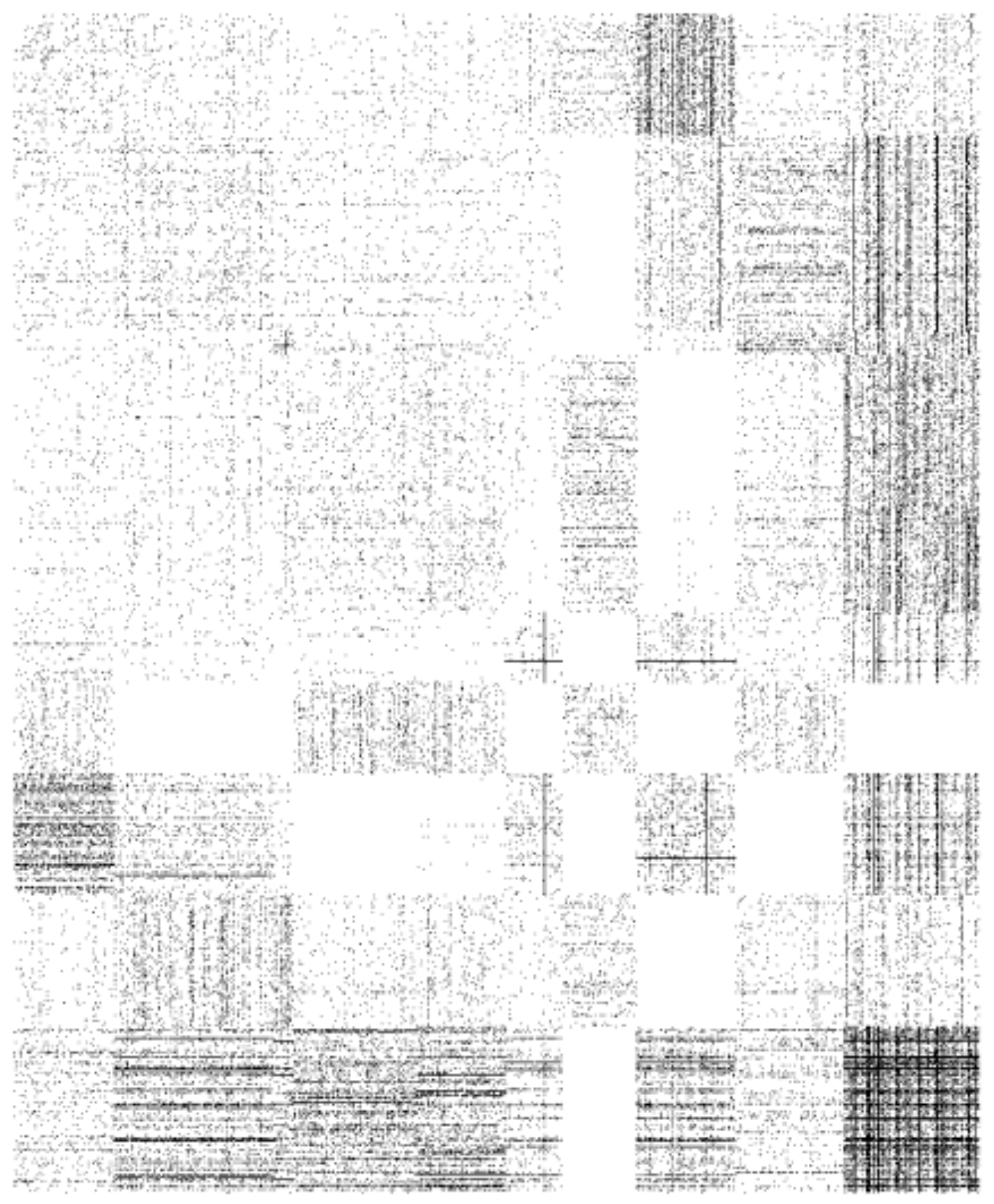}} \hspace{5mm}
\subfloat[]{\includegraphics[height=\hei]{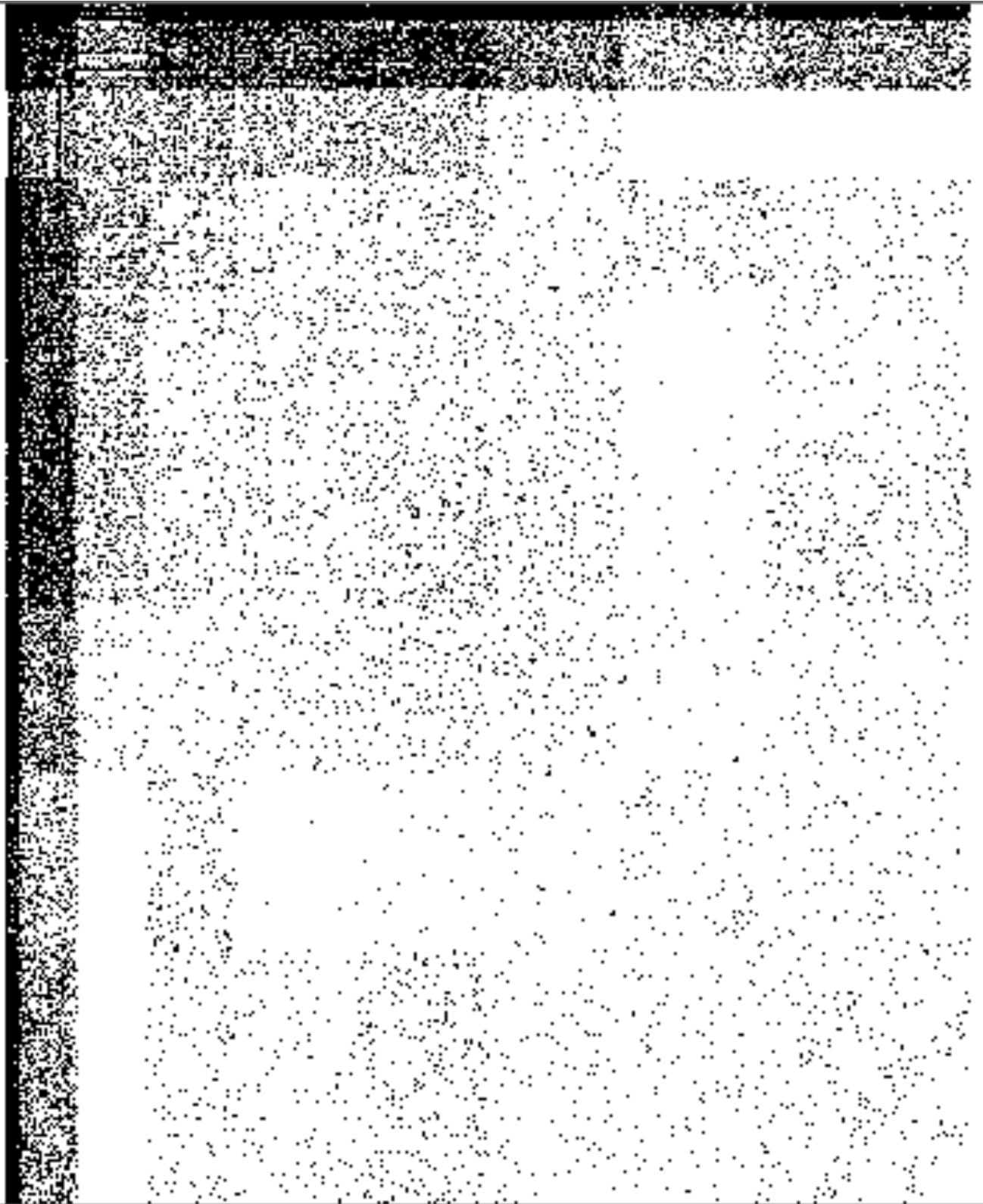}}
\vspace{10mm}
\caption{\label{asadj} The adjacency matrix of the AS graph ($n=22963$) with rows and columns organised according to: (a) raw data, (b) communities identified by distance matrix, (c) communities identified by distance matrix and degrees.}
\end{figure}

\begin{figure}[!htbp]
\centering
\setlength{\wid}{.33\textwidth}
\subfloat[]{\raisebox{0.2mm}{\includegraphics[width=\wid]{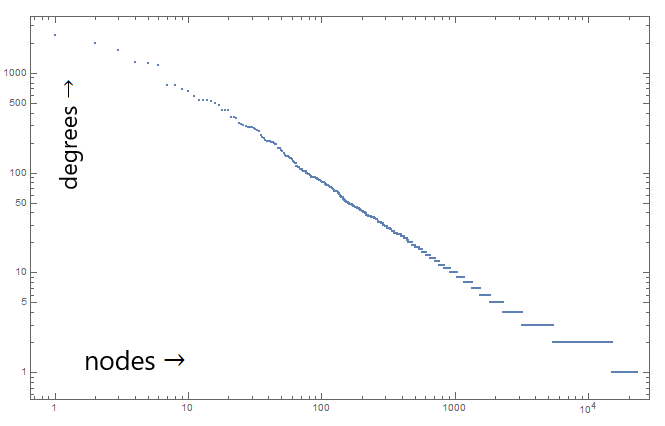}}}%
\subfloat[]{\raisebox{0.8mm}{\includegraphics[width=\wid]{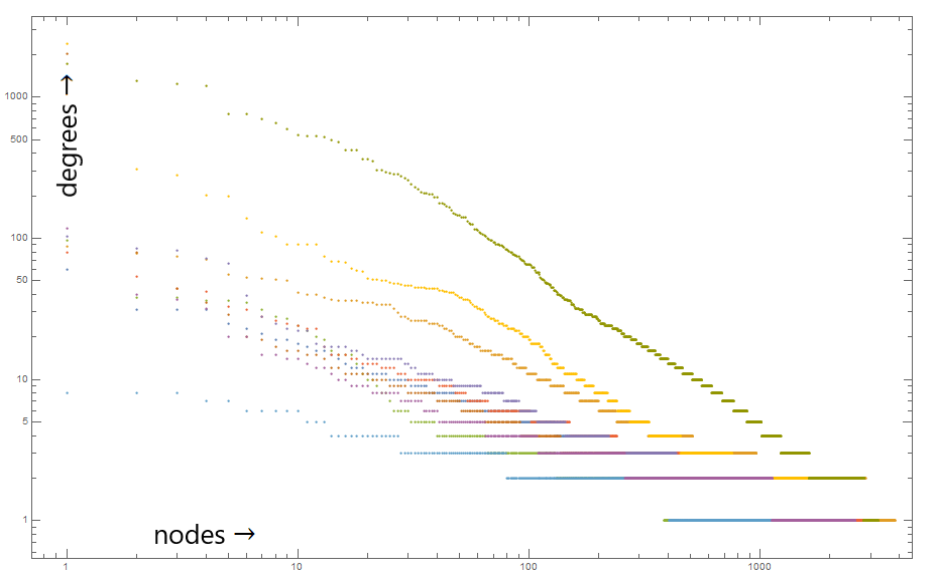}}}%
\subfloat[]{\includegraphics[width=\wid]{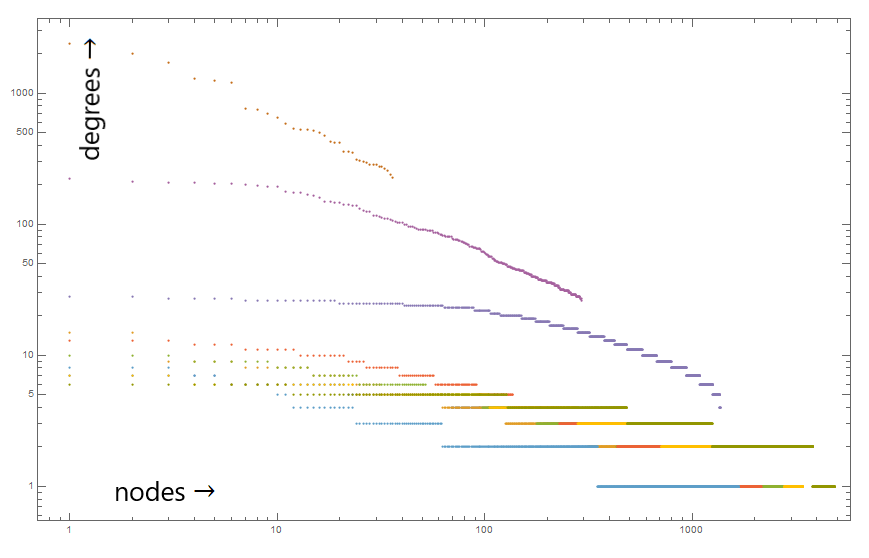}}%
\vspace{15mm}
\caption{\label{asdeg}
Degrees of nodes in the AS graph ($n=22963$) sorted from largest to smallest within each community (indicated by colour). (a) Full graph viewed as one community. (b) Nodes organised into communities identified by distance matrix. (c) Nodes organised into communities identified by distance matrix and degrees.
}
\end{figure}

We analyse the topology of the Internet by investigating a snapshot of the autonomous systems (AS) graph in 2006, reconstructed by Mark Newman from data collected by University of Oregon's Route Views Project\footnote{\url{https://github.com/gephi/gephi/wiki/Datasets}\label{DataAS}}. The graph has 22963 nodes (AS) and 48436 edges (neighbouring AS pairs). The adjacency matrix and the degrees are plotted in Fig.~\ref{asadj}:(a) and
Fig.~\ref{asdeg}:(a), respectively. The latter demonstrates an approximate power-law structure: six nodes have degree larger than 1000, whereas most nodes have degree less than 10. The graph has a soft hierarchical structure\footnote{\url{https://en.wikipedia.org/wiki/Tier_1_network}\label{DataWiki}} with the most important nodes contained in tier~1, the second most important nodes in tier~2, and so on. 

\subsection{Regular decomposition using distances}
\label{sec:ASDistances}

We partition the AS graph into communities by Algorithm~\ref{alg:RD} using $100$ randomly sampled columns of the distance matrix as data matrix. This appears sufficient for this type of network where we expect that the distances from a typical reference node depend heavily on the position of the node in the network hierarchy.  Using the method in Sec.~\ref{sec:CommunityCount}, we found that the most informative number of communities is $k=10$.
%
%
Fig.~\ref{asadj}:(b) displays the adjacency matrix of the AS graph organised by the identified community structure, showing that the communities are all rather large and of comparable size. The link densities inside and between communities are all low, and comparable to the overall link density. The degrees of nodes grouped into communities are displayed in Fig.~\ref{asdeg}:(b).
%
%
%



The results for the AS graph have similarities with the synthetic power-law graph in Sec.~\ref{sec:PowerLawDistances}. For instance, although all high-degree nodes are in the same community, this community also contains many low-degree nodes and thus has a low internal density. This can be seen in Fig.~\ref{asdeg}:(b) where the degree sequences of different communities overlap.  As a result, using only graph distances as the data matrix, we were not able to identify the tiers of the AS graph. For instance, the smallest community has much more nodes than there are tier-1 nodes\footref{DataWiki}.

\subsection{Regular decomposition using distances and degrees}

We repeat the community identification experiment of the AS graph by augmenting the 100-column data matrix used in Sec.~\ref{sec:ASDistances} with 1 column containing the node degrees. Again, $k=10$ is identified as the most informative number of communities. Fig.~\ref{asadj}:(c) displays the adjacency matrix of the graph organised according to the identified communities. The resulting community structure substantially differs from the one in the previous section. The smallest two communities in  Fig.~\ref{asadj}:(c) are dense and approximately correspond to tier~1 and tier~2 subnetworks of the AS graph. The degree sequences of the communities are shown in Fig.~\ref{asdeg}:(c).  There are three rather small and dense communities which contain all high-degree nodes, but no low-degree nodes.

\begin{figure}[!htbp]
\centering
\includegraphics[width=110mm]{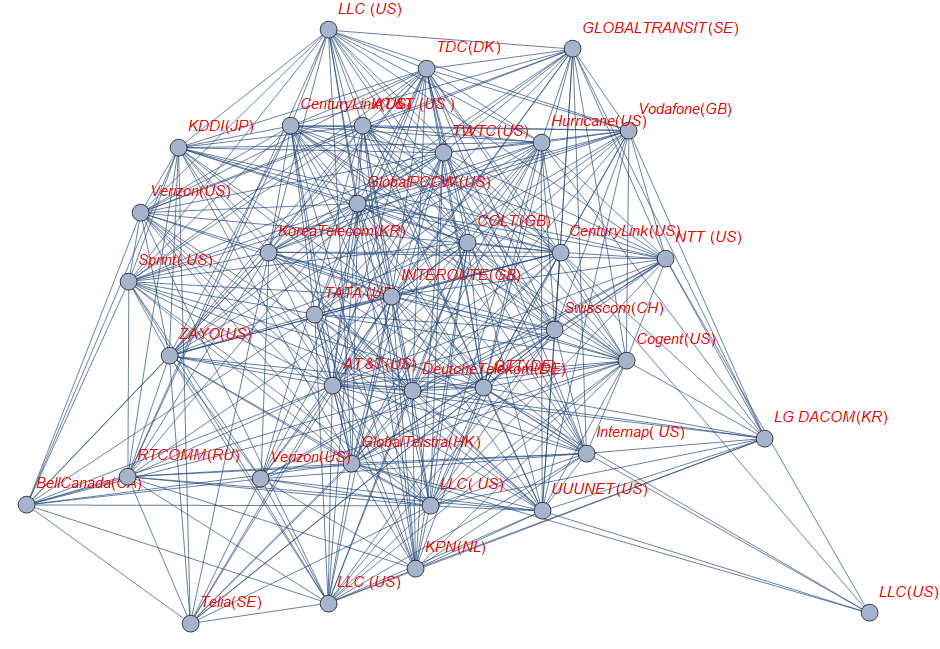}
\vspace{3mm}
\caption{\label{tier1} Subgraph of the AS graph induced by the smallest community identified by regular decomposition using distances and degrees.}
\end{figure}

\begin{figure}[!htbp]
\centering
\setlength{\wid}{.45\textwidth}
\subfloat[]{\raisebox{0.0mm}{\includegraphics[width=\wid]{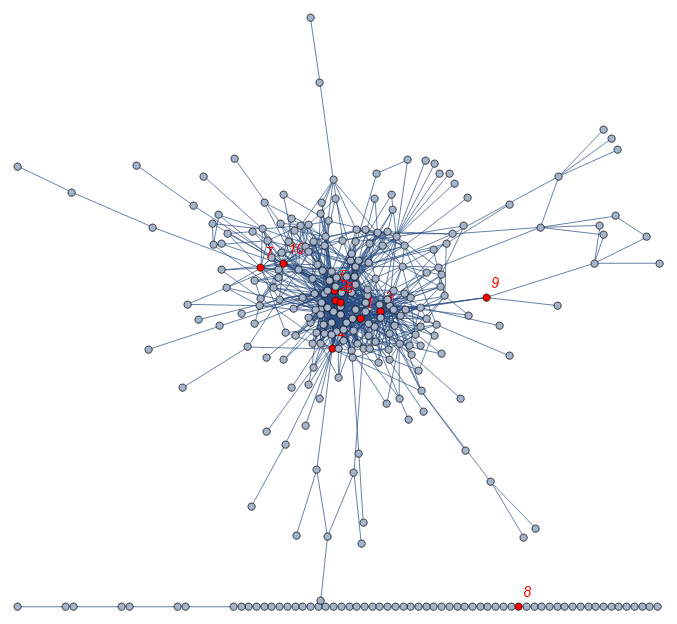}}} \hspace{5mm}
\subfloat[]{\raisebox{0.0mm}{\includegraphics[width=\wid]{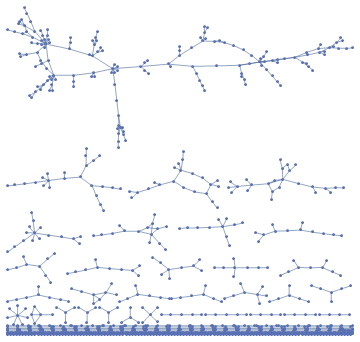}}}%
\vspace{17mm}
\caption{\label{tier2} Subgraphs of the AS graph induced by the second (a) and third (b) smallest communities identified by regular decomposition using distances and degrees. Ten nodes of highest overall degree in the second smallest community are highlighted in red: 1. Net Access Corporation, 2. Microsoft, 3. LondonInterconnectionPoint, 4. BT, 5. Internet Initiative Japan, 6. FrontierCommunicationsofAmerica, 7. MCICommunicationsServices, 8. INAP, 9. TransTeleCom, and 10. Telstra.}
\end{figure}

To assess the quality of the identified communities, we determined the AS identities in the three smallest and densest communities using a list of AS networks\footref{DataWiki} and an AS lookup tool\footnote{\url{https://www.bigdatacloud.com/asn-lookup}\label{DataASNames}}.
Our aim is to verify that the three identified communities are close to tier~1--3 subnetworks and contain the most important autonomous systems.  The smallest identified community (Fig.~\ref{tier1}) contains $36$ nodes, out of which
$23$ are tier~1 and the rest are tier~2. The members of tier~2 are important telecom carriers.  For example, ''Hurricane Electric''\footref{DataWiki} has a very high degree ($7061$), which explains why it is included in the smallest community by the regular decomposition algorithm. The discrepancy between tier~1 and the smallest identified community may be due to the that the AS graph topology deviates from a simple power-law graph \citep[e.g.][]{willinger}.
The second smallest identified community is shown in Fig.~\ref{tier2}:(a). The largest degrees are around $200$. In this community, the ten nodes of highest overall degree do not belong to tier~$1$, but nevertheless correspond to networks operated by major companies such as British Telecom (UK) and Microsoft (US). The third smallest community, displayed in Fig.~\ref{tier2}:(b), is much sparser and its highest degrees are around $20$. Top-degree members are Telefonica Data S.A. (BR), Orano (US), and Harvard University (US). 

We conclude that augmenting the graph distance matrix by a column containing the node degrees allows to identify much more meaningful communities, compared to only using the distance matrix. The regular decomposition method was able to
identify central carriers in the top tiers with good accuracy from a large data set. In particular, we discovered a dense tier 1-rich subnetwork. The suggested method could be used even for extremely large graphs encountered in areas such as biology and social networks, where it might be impossible to acquire the entire graph for analysis. Our methods need only a limited sample of shortest paths between a set of sampled nodes, and node degrees. Community detection on such a sample results in a model which can be used to classify any other node outside the sample. It has the potential to rapidly detect soft hierarchies in massive networks.

\section{World airline graph}
\label{sec:AirlineNetwork}

\begin{figure}[!htbp]
\centering
\includegraphics[width=117mm]{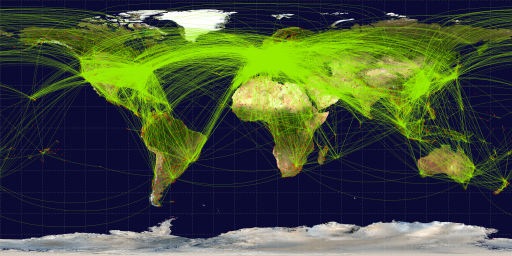}
\vspace{3mm}
\caption{\label{airlines} Geographic projection of the world airline graph.}
\end{figure}

\begin{figure}[!htbp]
\centering
\includegraphics[width=120mm]{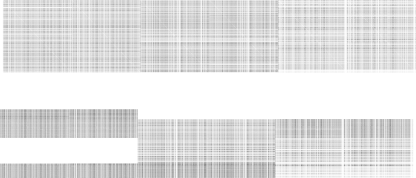}
\vspace{15mm}
\caption{\label{d} Top: Grey-scale visualisation of the data matrix of the world airline graph. The white elements correspond to undefined distances. The matrix has $691$ rows (airports) and $4146$ columns (directed graph distances in three layers). Each layer occupies a band of columns of equal width and is roughly visible in the picture. Bottom: The same data matrix with rows reorganised into $6$ communities identified by regular decomposition using layerwise distances}
\end{figure}

We demonstrate how the regular decomposition method can be applied to a directed multilayer graph defined as a 
collection of $s$ graphs with a common set of $n$ nodes, each graph representing one layer.
As a concrete example, we used a world airline graph\footnote{\url{https://openflights.org/data.html}\label{DataFlight}}
consisting of 3321 nodes (airports), 67663 links (flights), and 548 layers (airlines) displayed in Fig.~\ref{airlines}.  We extracted the three largest airlines (American Airlines, United Airlines, Air France) in June 2014, resulting in a directed multilayer graph with $n=691$ nodes and $s=3$ layers.
Fig.~\ref{d} displays the data matrix corresponding to the concatenation of three directed graph distance matrices and their transposes (see Eq.~\ref{eq:MultilayerDirected}).

\subsection{Regular decomposition using layerwise distances}
\label{sec:AirlineRD}

Using Algorithm~\ref{alg:RD} and the method in Sec.~\ref{sec:CommunityCount}, we discovered $k=6$ as the most informative number of communities, and identified the corresponding communities. The resulting data matrix, organised by the identified communities is shown at the bottom of Fig.~\ref{d}. The smallest community has 9 nodes consisting of airports in the Middle East (4), Europe (2), East Asia (1), and South America (1).
The second smallest community has 13 nodes consisting of airports in East Asia (5), Africa (3), South-East Asia (2), Pacific (2), North-America (1). The remaining four communities are of comparable size: One of them has 27 airports in France and 85 in USA, and the other three have most airports located in the USA. 

\begin{figure}[!htbp]
\centering
\subfloat[]{\includegraphics[height=50mm]{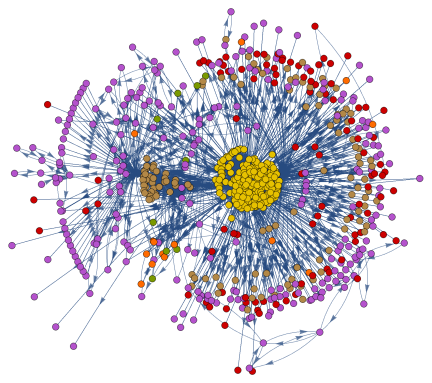}} \hspace{2mm}
\subfloat[]{\raisebox{3mm}{\includegraphics[height=45mm]{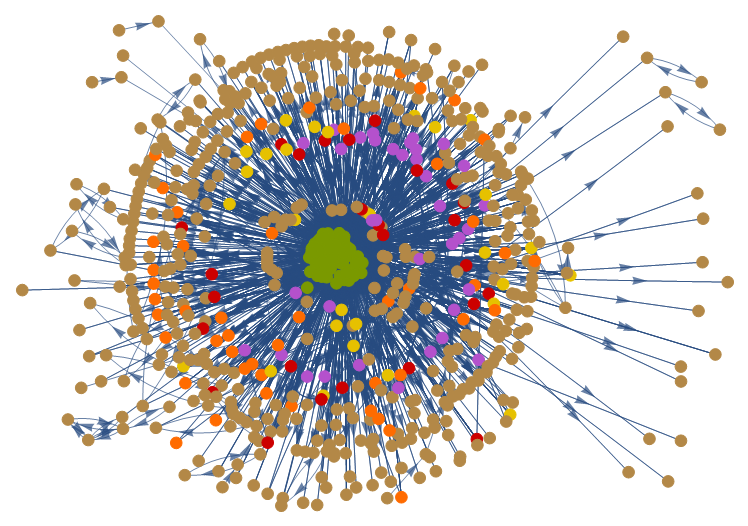}}}
\vspace{8mm}
\caption{\label{aircommunities} Communities of the world airline graph identified by (a) regular decomposition using layerwise distances, (b) SBM fitting with a layer-aggregated adjacency matrix.}
\end{figure}

\section{Comparison with other community detection and data compression methods}
In this Section we make a quantitative assessment of communities found by regular decomposition with respect to some other community detection and data analysis methods. As the test cases we use the real-life networks analysed in the previous Sections.  
\subsection{Community detection}

\subsubsection{Internet autonomous systems graph}

As stated in the introduction, our aim is to have a community detection method which identifies communities which reflect various aspects of data associated with the network, and their role in the graph topology. There are of course many existing powerful community detection methods which can do this in some particular cases.  We illustrate this point by finding communities in the AS graph (Sec.~\ref{sec:AS}) using two popular methods: modularity maximisation and stochastic block model fitting.

\begin{figure}[!htbp]
\centering
\subfloat[]{\includegraphics[width=65mm]{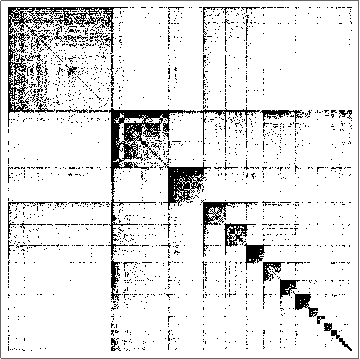}} \hspace{3mm}
\subfloat[]{\includegraphics[width=65mm]{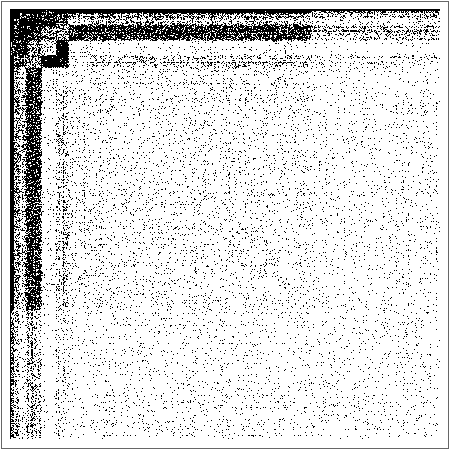}}
\vspace{8mm}
\caption{\label{modularity} Adjacency matrix of the AS graph organised according to (a) 25 communities identified by modularity maximisation, (b) 10 communities identified by SBM fitting.}
\end{figure}

\begin{figure}[!htbp]
\centering
\includegraphics[width=100mm]{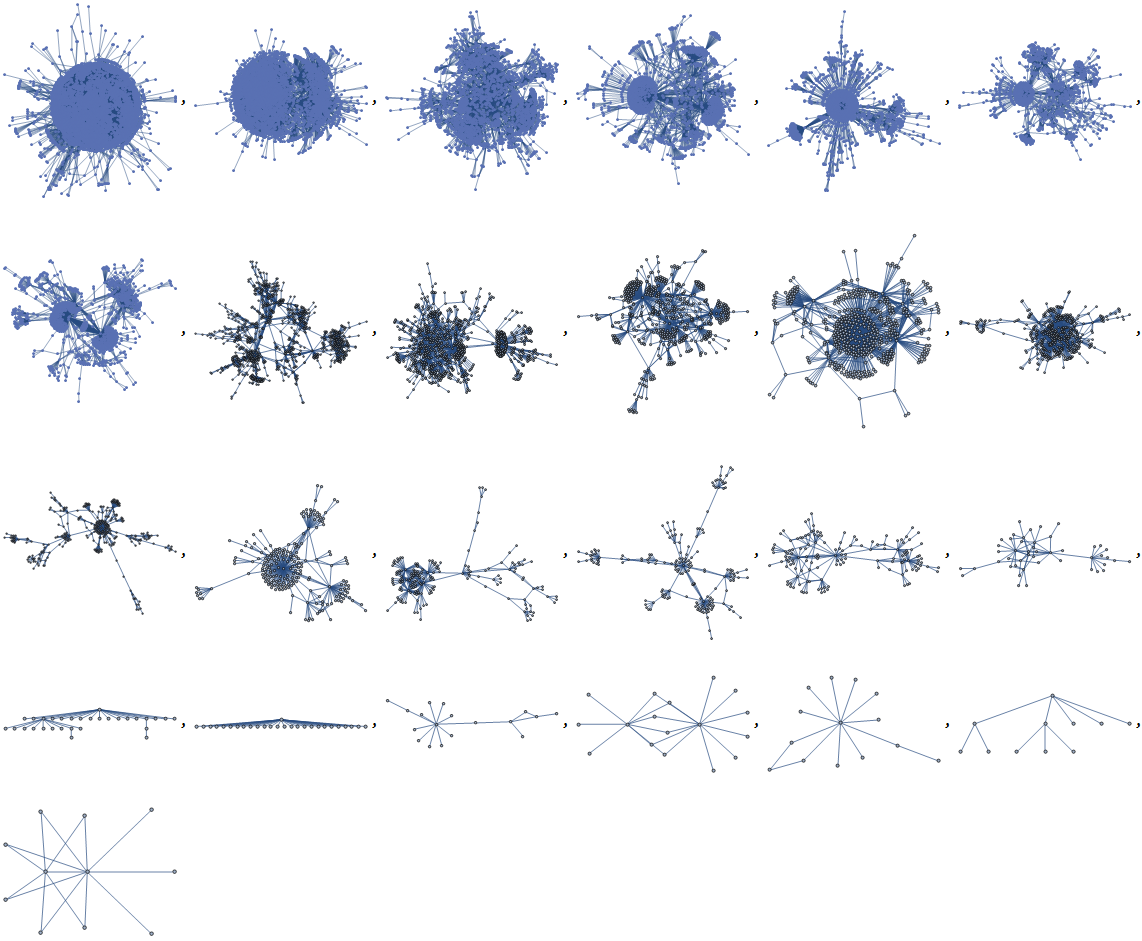}
\vspace{5mm}
\caption{\label{modularitysub} Subgraphs of the AS graph induced by the $25$ communities 
identified by modularity maximisation.}
\end{figure}

Modularity maximisation \citep{modularity} aims to find densely connected communities which have as little as possible links between the communities. 
Fig.~\ref{modularity}:(a) displays the adjacency matrix organised according to the 25 identified communities, and Fig.~\ref{modularitysub} illustrates the subgraphs induced by the communities. 
As expected, the community structure determined by modularity maximisation is substantially different from the community structures identified by regular decomposition in Fig.~\ref{asadj}:(b--c). The subgraphs in Fig.~\ref{modularitysub} do not respect the tiered structure found with regular decomposition. For instance, the high-degree nodes forming tier~1 are embedded in very large communities, which can be inferred from Fig.~\ref{modularitysub}.

\begin{figure}[!htbp]
\centering
\includegraphics[width=80mm]{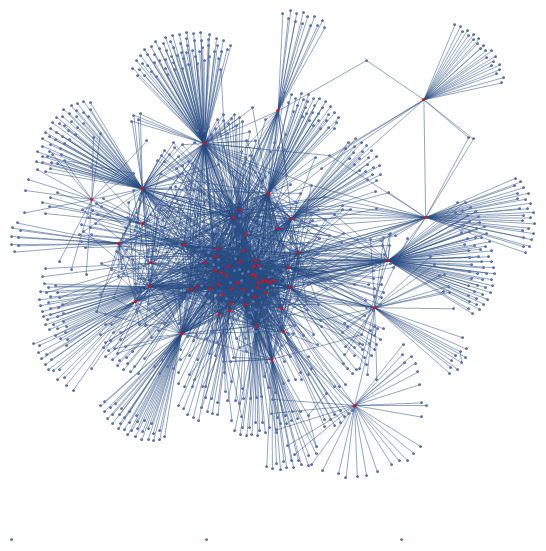}
\caption{\label{pair} A subgraph of the AS graph induced by two distinct communities (red, blue) identified by SBM fitting. The nodes of the blue community are almost entirely low-degree neighbours of the nodes in the red community.}
\end{figure}

Stochastic block model (SBM) fitting \citep{yunpeng} uses the adjacency matrix to find communities in which relations inside and between the communities are like those in classical random graphs. Information-theoretic model fitting can be used to find the communities. We used the regular decomposition method for this. 
For computational tractability, instead of the full adjacency matrix we restricted to the largest connected component of the subgraph induced by a uniform random sample of 10000 nodes.
After identifying the communities of the restricted graph, the parameters of the model were used to classify all nodes of the graph. A smaller sample is not feasible, because the resulting induced subgraph would hardly have any links. This is in stark contrast to regular decomposition using graph distances, where a sample of 100 nodes suffices. Fig.~\ref{modularity}:(b) shows the resulting community structure, where a dense tier-1-like community appears, but a deeper tier hierarchy seems quite weakly represented, as manifested by a large unstructured block of low-degree nodes. In contrast, a much finer community structure is identified by regular decomposition based on distances and degrees in Fig.~\ref{asadj}:(c).  A typical subgraph induced by a pair of communities identified by SBM fitting is a well-connected graph in which one part is like a neighbourhood of the other community, see Fig.~\ref{pair}. On the other hand, in a community structure identified by distance profiles, such pairs are typically not connected --- this is simply because many distances are larger than one.  

\begin{figure}[!htbp]
\centering
\setlength{\wid}{.33\textwidth}
\subfloat[]{\includegraphics[width=\wid]{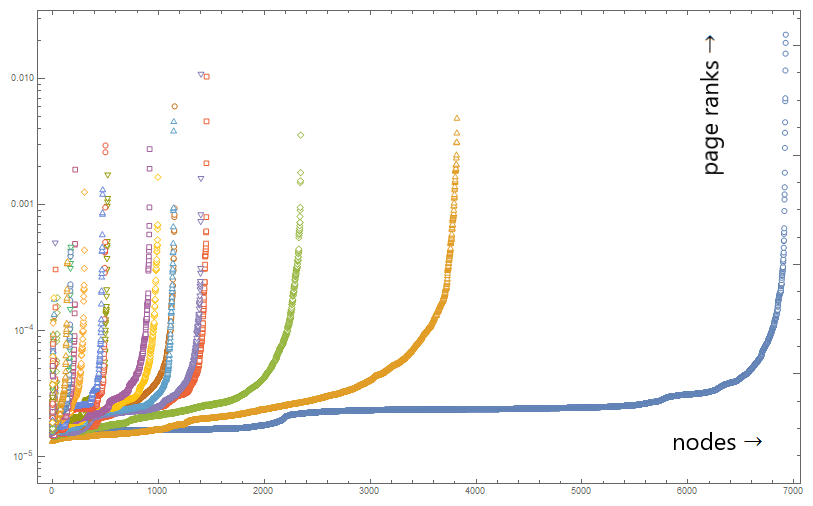}}%
\subfloat[]{\includegraphics[width=\wid]{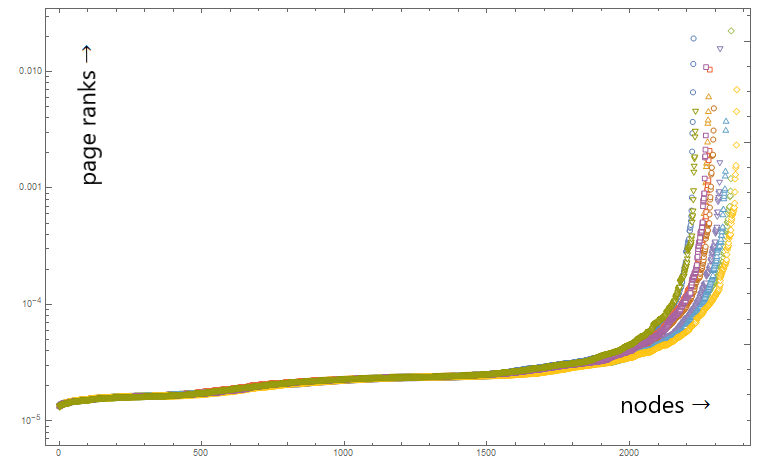}}%
\subfloat[]{\includegraphics[width=\wid]{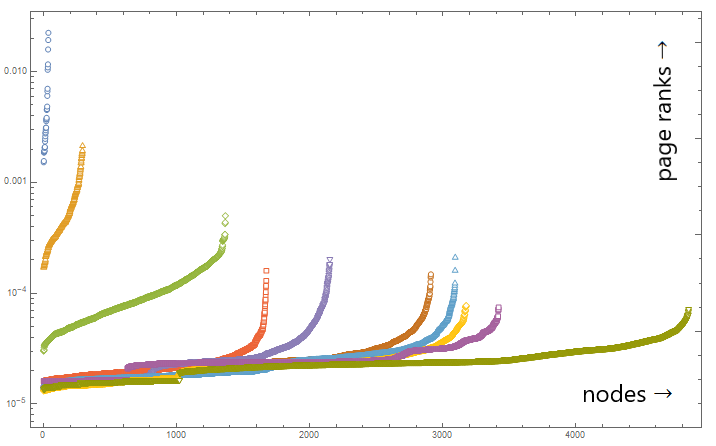}}%
\vspace{15mm}
\caption{\label{pagerank}
PageRank centrality of AS graph nodes in communities identified using three alternative methods, and sorted from smallest to largest within each community (indicated by colour). (a) Modularity maximisation, (b) RD using distances, (c) RD using distances and degrees. Only the last method can group the most central nodes into separate communities, indicated by blue and orange circles in (c).}
\end{figure}

For a more quantitative comparison, we computed the PageRank centrality of the network nodes (with damping factor 0.8), and plotted in Fig.~\ref{pagerank} the PageRank distributions within communities identified by three methods: modularity maximisation, RD using distances, and RD using distances and degrees.
We see that the latter is the only method able to separate nodes of high PageRank into a common community.
This illustrates the ability of our method in identifying community structures associated with the topological roles of nodes in the network.

\subsubsection{World airline graph}

\begin{figure}[!htbp]
\centering
\subfloat[]{\includegraphics[width=60mm]{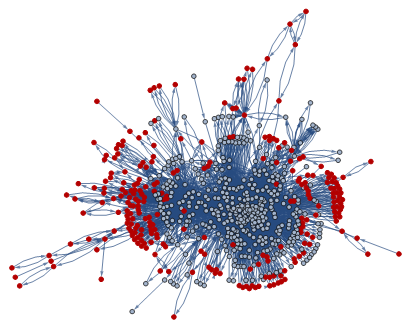}}
\subfloat[]{\includegraphics[width=60mm]{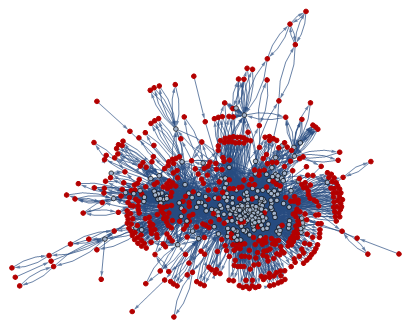}} \hspace{5mm}
\vspace{10mm}
\caption{\label{largecom} 
Largest community (highlighted in red) of the world airline graph identified using (a) regular decomposition using layerwise distances, (b) SBM fitting with a layer-aggregated adjacency matrix.}
\end{figure}

\begin{figure}[!htbp]
\centering
\includegraphics[width=60mm]{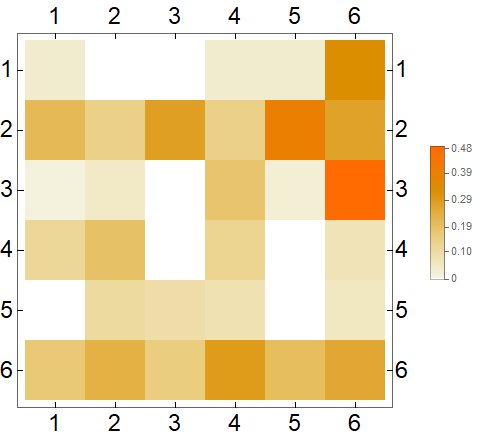}
\vspace{18mm}
\caption{\label{overlap} Overlap of communities determined by SBM fitting using layer-aggregated graph (row index) and by regular decomposition using layerwise distances (column index). The $(i,j)$-entry is computed as the number of common nodes in community pair $(i,j)$ divided by the number of nodes in the larger community in the pair. Only the pair $(3,6)$ indicates a substantial value around $0.5$. This pair corresponds to the peripheral communities highlighted in Fig.\ref{largecom}.}
\end{figure}

We compare the community structure of the world airline graph determined in Sec.~\ref{sec:AirlineRD} with a more customary SBM fitting applied to the adjacency matrix of the undirected graph obtained by collapsing the layers and ignoring link directions. We impose the same number $k=6$ of communities as previously. The resulting community structure, visualised in Fig.~\ref{aircommunities}:(b), bears some similarity with the one in Fig.~\ref{aircommunities}:(a), but mostly in the periphery of the network which forms the largest community in both cases. The peripheral communities are highlighted in Fig.~\ref{largecom}. Furthermore, Fig.~\ref{overlap} displays the overlap matrix of communities. The overlap is weak and the community structures are quite different. If the community structures were similar, there should be at each row exactly one strong element in this matrix and those strong elements would be all in different columns.

\subsection{Data compression}

%
%

%

%

A natural measure of our method is the compression rate of the data matrix used in community detection.  This experiment is done on the distance matrix for the AS graph. As the uncompressed description length, we used Mathematica's internal memory requirement for storing a distance matrix (ByteCount), which for the AS graph equals $L_0 \approx 1.7 \times 10^{10}$ bits. First, as a standard compression, we applied Mathematica's built-in implementation of the zlib\footnote{\url{https://zlib.net}} algorithm.  
Second, we computed the description length $L$ for the distance matrix by applying RD.  This consists of the Shannon code length, which is minus base-2 logarithm of the probability of the distance matrix in the probabilistic model induced by the community structure, and the code lengths of the parameters.  We used the leading part of the prefix code lengths of integers \citep{grunwald}. For a positive integer $m$, the code length of the integer is
$\ceiling{\log_2 m} + \ceiling{\log_2\log_2 m}+\cdots$, in which $\log_2$ is iterated as long as the result remains positive, after which the sum is truncated. The parameters we need to encode are the expectations of $k\times n$ Poisson variables, $k=10$, and $n=22963$, and the partition into communities which can be represented as an $n$-vector with coordinates in $\{1,\dots,10\}$. Third, we computed similar code length for community structure found using node degrees along with the distance matrix. The compression ratio $L/L_0$ is displayed in Table~\ref{tab:table1}. We see that zlib compresses the original data by around $9$ times while RD using distances does a better job with around $12$ times compression. Augmenting the RD by also using degrees leads to a slight further improvement.

We also repeated this experiment for the world airline graph for which the uncompressed layerwise distance matrix requires $L_0 \approx 5.5 \times 10^8$ bits. The zlib algorithm compresses this data by around $80$ times, while regular decomposition on the layerwise distance matrix is able to compress $235$ times, almost three times more than zlib, see Table~\ref{tab:table1}.

\begin{table}[h!]
\centering
\caption{Compression ratio for the AS graph and the world airline graph using three methods: zlib compression algorithm, RD using distances, and RD using distances and degrees.\label{tab:table1}}
\begin{tabular}{lrrr}
 \hline
 & AS graph & Airline graph \\
 \hline
 zlib & 9.2 & 80.4 \\
 RD using distances & 12.3 & 235.6 \\
 RD using distances and degrees & 12.6 & -- \\
 \hline
\end{tabular}
\end{table}

\section{Conclusion}

We demonstrated a unified approach of finding network communities in large and sparse multilayer graphs, based on extending the regular decomposition method to handle data matrices with an arbitrary number of columns representing various types of data associated with nodes. We demonstrated our method by analysing graph distance matrices augmented with a column of node degrees. Our method has a low computational complexity allowing to handle massive input graphs, and it 
also tolerates missing data entries.  We illustrated the method with a synthetic power-law graph and two real graphs: a snapshot of the Internet topology and a directed multilayer graph describing the world airline topology.  In the latter case, as data matrix we used a concatenation of graph distance matrices from different layers, which allowed to find meaningful communities despite massive amounts of missing data.
In contrast to popular community detection methods, such as modularity maximisation and stochastic block model fitting, our method appears better suited for identifying community structures aligned with tiered hierarchies often encountered in scale-free complex networks. 
When applied to distance matrices, our method implicitly assumes that graph distances are Poisson distributed and mutually independent. 
This assumption was motivated by computational tractability instead of a good fit to data.  However, because graph distances in scale-free networks are known to be highly concentrated around their mean even for heavy-tailed degree distributions \citep{VanDerHofstad_Hooghiemstra_Znamenski_2007,VanDerHoorn_Olvera-Cravioto_2018}, the Poisson assumption may not be overly unrealistic.  For carrying out a theoretical analysis of consistency of our method applied to distance matrices, an important future problem is to first analyse joint distributions of distances in degree-corrected stochastic block models, extending state-of-the-art result obtained for random graphs without communities \citep{Bhamidi_VanDerHofstad_Hooghiemstra_2017,Jorritsma_Komjathy_2020}.
Although the experiments carried out in this work were restricted to topological data matrices that can be deduced from the graph adjacency matrix, the regular decomposition method allows to incorporate an arbitrary number of auxiliary columns to the data matrix.  This opens up ways to analyse and model network problems in which nontopological node-associated data plays an important role in forming graph communities, and remains an important problem of further study.

{\small
\paragraph{Acknowledgments.}
We thank Remco van der Hofstad for illuminating discussions, Mark Newman for commenting the AS graph data set, and anonymous referees for helpful remarks on improving the presentation. We thank  the action editor and anonymous referees and for helpful remarks and suggestions.  VTT’s part of this work was supported by BusinessFinland - Real-Time AI-Supported Ore Grade Evaluation for Automated Mining (RAGE) and Quantum Technologies Industrial (QuTi) projects.
}
 
\bibliographystyle{apalike}
\bibliography{NWS_Revision2.bib}
\end{document}